\documentclass[a4paper,11pt]{article}
\pdfoutput=1 

\usepackage{jinstpub} 
\usepackage{lineno}

\title{\boldmath Performance evaluation of a silicon strip detector for positrons/electrons from a pulsed a muon beam}

\author[a]{T.~Aoyagi}
\author[a]{Y.~Honda}
\author[b]{H.~Ikeda}
\author[c]{M.~Ikeno}
\author[f,g]{K.~Kawagoe}
\author[c]{T.~Kohriki}
\author[e]{T.~Kume}
\author[c]{T.~Mibe}
\author[a]{K.~Namba}
\author[d]{S.~Nishimura}
\author[c]{N.~Saito}
\author[c]{O.~Sasaki}
\author[e]{N.~Sato}
\author[c]{Y.~Sato}
\author[c]{H.~Sendai}
\author[d]{K.~Shimomura}
\author[f,h]{S.~Shirabe}
\author[c]{M.~Shoji}
\author[a]{T.~Suda}
\author[f]{T.~Suehara}
\author[e]{T.~Takatomi}
\author[c]{M.~Tanaka}
\author[f]{J.~Tojo}
\author[a]{K.~Tsukada}
\author[c]{T.~Uchida}
\author[i]{T.~Ushizawa}
\author[a]{H.~Wauke}
\author[g,1]{T.~Yamanaka}
\author[g]{T.~Yoshioka}

\affiliation[a]{Research Center for Electron Photon Science, Tohoku University,\\
1-2-1 Mikamine Taihaku-ku, Sendai, Japan}
\affiliation[b]{Institute of Space and Astronautical Science, Japan Aerospace Exploration Agency,\\
3-1-1 Yoshinodai, Sagamihara, Japan}
\affiliation[c]{Institute of Particle and Nuclear Studies, High Energy Accelerator Research Organization,\\
1-1 Oho, Tsukuba, Japan}
\affiliation[d]{Institute of Materials Structure Science, High Energy Accelerator Research Organization,\\
1-1 Oho, Tsukuba, Japan}
\affiliation[e]{Mechanical Engineering Center, High Energy Accelerator Research Organization,\\
1-1 Oho, Tsukuba, Japan}
\affiliation[f]{Department of Physics, Kyushu University,\\
744 Motooka Nishi-ku, Fukuoka, Japan}
\affiliation[g]{Research Center for Advanced Particle Physics, Kyushu University,\\
744 Motooka Nishi-ku, Fukuoka, Japan}
\affiliation[h]{Department of Physics, Tokyo Institute of Technology,\\
2-12-1 Ookayama Meguro-ku, Tokyo, Japan}
\affiliation[i]{Department of Particle and Nuclear Physics, Graduate University for Advanced Studies,\\
1-1 Oho, Tsukuba, Japan}

\emailAdd{yamanaka\_at\_artsci.kyushu-u.ac.jp}

\abstract{
A high-intensity pulsed muon beam is becoming available
at the at the Japan Proton Accelerator Research Complex (J-PARC).
Many experiments to study fundamental physics using this high-intensity
muon beam are proposed. An experiment to measure the muon magnetic moment anomaly ($g-2$)
and the muon electric dipole moment (EDM) is one of 
these experiments and it requires a tracking detector for positrons
from muon decay. Fine segmentation is required in a detector to tolerate the high rate
of positrons. The time resolution is required to be much better than the muon anomalous spin precession
period while a buffer depth of a front-end electronics needs to be much longer than the accelerated muon lifetime.
Requirements of this detector also meet requirements of
a measurement of the muonium
hyperfine structure interval at the J-PARC and another experiment
to measure the proton charge radius at Tohoku University.
We have developed a single-sided silicon strip sensor with a 190~$\mu$m pitch, a
front-end electronics with a sampling rate of 200~MHz and a buffer memory depth of 8192,
and a data acquisition system based on 
DAQ-Middleware for the J-PARC muon $g-2$/EDM experiment.
We have fabricated detector modules consisting of this sensor and the front-end
electronics.
Performance of fabricated detector modules was evaluated at a laboratory and
a beam test using the positron beam at Tohoku University.
The detector is confirmed to satisfy all requirements of the experiments
except for the time walk,
which will be solved by the next version of a front-end electronics.
}


\arxivnumber{1910.13087} 

\begin{document}
\maketitle
\flushbottom

\section{Introduction}
\label{sec:introduction}

\subsection{Muon research at J-PARC}

A high-intensity pulsed muon beam at the order
of $10^7$~muons/s on average 
is becoming available at the Muon Science Establishment (MUSE) located at 
the Material and
Life Science Experimental Facility (MLF) in the Japan Proton
Accelerator Research Complex (J-PARC)~\cite{MuSE}.
The repetition rate of the beam is 25~Hz. The beam spill consists of two bunches
separated by 600~ns with a bunch of 100-150~ns full width. 
 The average intensity
will be of the order of $10^8$~muons/s when the 3~GeV proton
synchrotron ring of the J-PARC is fully operated
and the proton beam power of 1~MW is reached as designed, which is
used for generating a muon beam.
Short time operation of the proton beam power of 1~MW was
 demonstrated in 2019 and it will be provided to beam users soon.
In a new muon beam line, the H-line, which is being constructed,
another factor of improvement is expected in the muon beam intensity~\cite{H-line}. 
Several experiments are planned at the H-line
assuming the beam intensity of $10^{8}$~muons/s.
A high-intensity muon beam is a key to search for new physics beyond the
Standard Model in precise measurements of fundamental parameters
of the muon with high statistics and 
is also useful for material science studies
based on muon spin spectroscopy.

\subsection{Muon $g-2$/EDM experiment at J-PARC}

A new experiment is proposed to measure the muon magnetic moment anomaly ($g-2$)
and the muon electric dipole moment (EDM) at the J-PARC MUSE H-line~\cite{E34_PTEP2019}.
The experiment uses a different method from the E821 experiment at Brookhaven National Laboratory~\cite{E821_PRD2006}
or the E989 experiment at Fermilab~\cite{E989_arxiv2015}
and aims to measure the muon $g-2$ with a statistical uncertainty
of 450 parts per billion and the muon EDM with a sensitivity of
$1.5\times 10^{-21}$~$e~\cdot$~cm.
In this experiment, a muon beam produced by a primary proton beam
is slowed down and thermalized in a material. The thermalized muon beam
is accelerated to 300~MeV/$c$ to produce a low-emittance muon beam.
This muon beam is injected into 
a storage ring using a novel 3D spiral injection method.
The storage ring is a 3~T magnetic resonance imaging-type solenoid
with high field uniformity for the muon storage region with an orbit
diameter of 66~cm.
A tracking detector is placed inside of the magnet and 
is required to reconstruct positron tracks from muon decay
and to estimate the muon decay time.
The anomalous spin precession frequency  
of muons in this magnetic field
is measured by the muon decay time distribution with higher-momentum positrons.
The sensitivity to the muon spin becomes maximum when positrons
with momenta above 200~MeV/$c$ are counted.


\subsection{Other experiments}


The MuSEUM experiment is another experiment proposed
at the J-PARC MUSE H-line to measure the muonium hyperfine structure
interval and the ratio of muon to proton magnetic moments
by using the muonium spin resonance method at high magnetic 
field~\cite{MuSEUM}.
The number of positrons from decay of the muon beam
after forming a muonium and experiencing a resonant microwave
is counted as a function of time in this
experiment. A detector with fine segementation and a high sampling rate is
required to tolerate positrons from a high-intensity muon beam
of $\sim4 \times 10^{6}$ muons in one spill.

Requirements for detectors in these experiments 
meet requirements of another experiment 
proposed at the Research Center for Electron Photon Science (ELPH),
Tohoku University,
which will measure the proton charge radius by low-energy
electron scattering at the lowest momentum ever achieved (ULQ2 experiment)~\cite{ULQ2}.
In this experiment, electrons in the energy range of 20-60~MeV
are scattered at a target, and the momentum and the scattering angle
of scattered electrons need to be measured.
An electron spectrometer with the momentum resolution of $8\times 10^{-4}$ and
the scattering angle resolution of 5~mrad is 
developed. To achieve this resolution, a detector
with a position resolution much better than 1~mm is
required at a focal plane. The expected rate of electrons at the focal plane
is 30~MHz in an area of $10\times 10$~cm$^{2}$.

\subsection{Requirements for tracking system}


A positron tracking detector for the J-PARC muon $g-2$/EDM experiment
is required to be sensitive in a volume with a radius of 290~mm
and a height of 400~mm. In this volume, 40 layers of detection planes
are required to reconstruct positron tracks with the momentum
above 200~MeV/$c$ even in the highest muon decay rate.
To cover this volume with a reasonable cost,
we choose silicon strip detector technology.
The maximum number of positrons from muon decay
is expected to be six per 1~ns and the maximum hit rate will reach 150~kHz/mm$^{2}$.
The detector is required to tolerate this hit rate and
to measure up to four minimum ionizing particle (MIP) charge
to accommodate pile-up of hits. 
The signal-to-noise ratio greater than 15 is required to reconstruct
positron tracks in high efficiency. 
Since the anomalous spin precession period in the magnetic field of 3~T is
about 2.1~$\mu$s, the time resolution is required to be much better than this 
period. On the other hand, the data needs to be stored sufficiently longer than
this period as well as the lifetime of muon with the momentum of 300~MeV/$c$ ($\sim$6.6~$\mu$s).
Another requirement comes from
the pile-up of detector hits. If there are pile-up hits in the same
sensor strip, a sensor signal waveform is distorted and the recorded
timing is shifted. 
Since the rate of pile-up changes during
the data taking period due to the lifetime of muon, it will bias
the spin precession frequency measurement. This effect can be
mitigated by a readout system with a small time walk. To constrain
this effect, the time walk is required to be less than 1~ns over a signal range
of 0.5-3 MIP charge.


To satisfy these requirements, we have developed a
single-sided silicon strip sensor and a front-end readout electronics
optimized for the experiment.
We have fabricated 
detector modules consisting of this sensor and the readout electronics
controlling their qualities.
Performance of fabricated detector modules was evaluated at a laboratory and
a beam test using the positron beam
at the ELPH, Tohoku University. The design and the performance
of the detector modules are presented in this article.

\section{Detector system}
\label{sec:detector_system}

The detector module consists of a silicon strip sensor and
two front-end readout boards.

\subsection{Sensor}

\begin{figure}[htbp]
  \begin{center}
    \includegraphics[height=0.28\textheight]{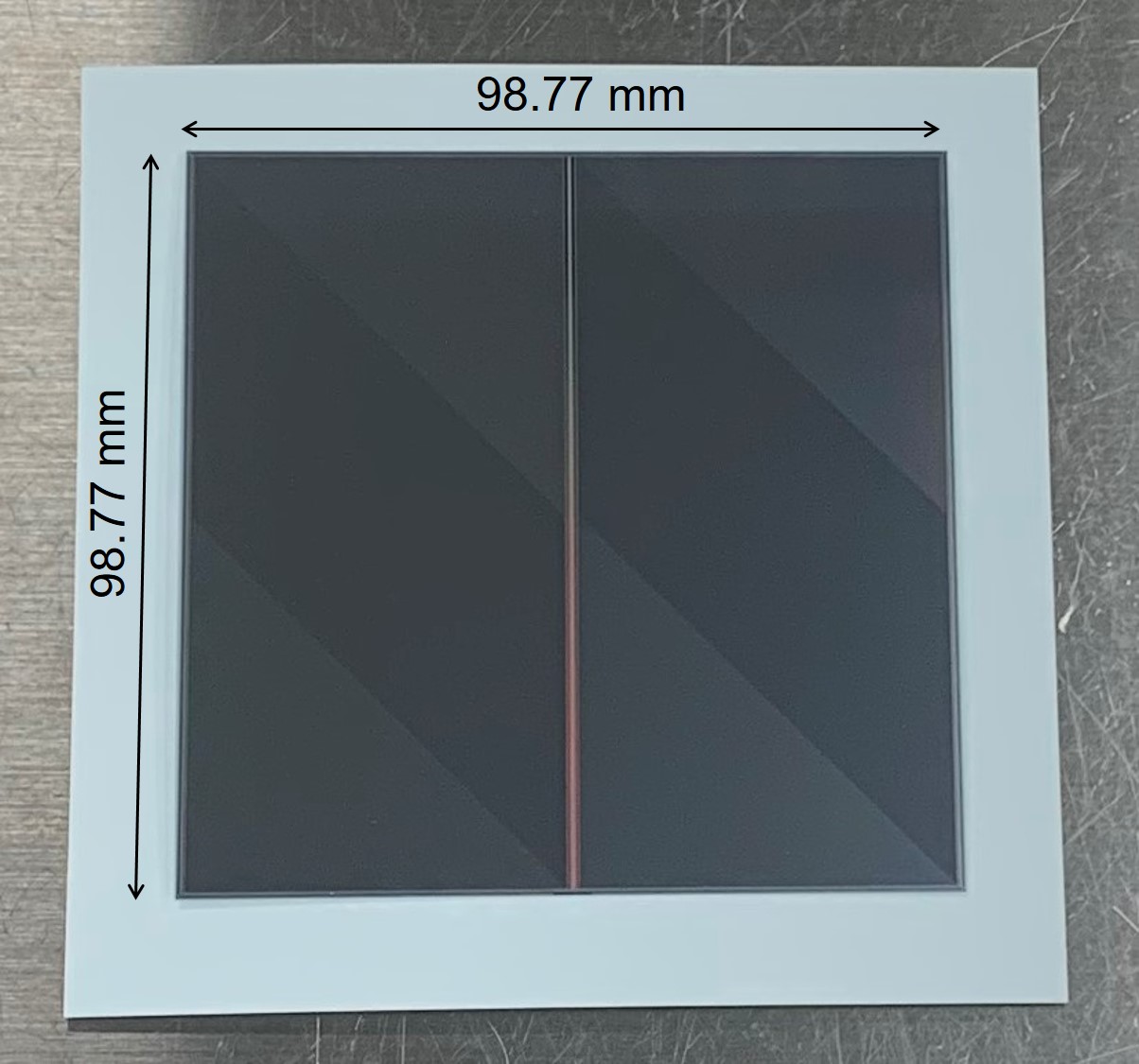}
    \includegraphics[height=0.28\textheight]{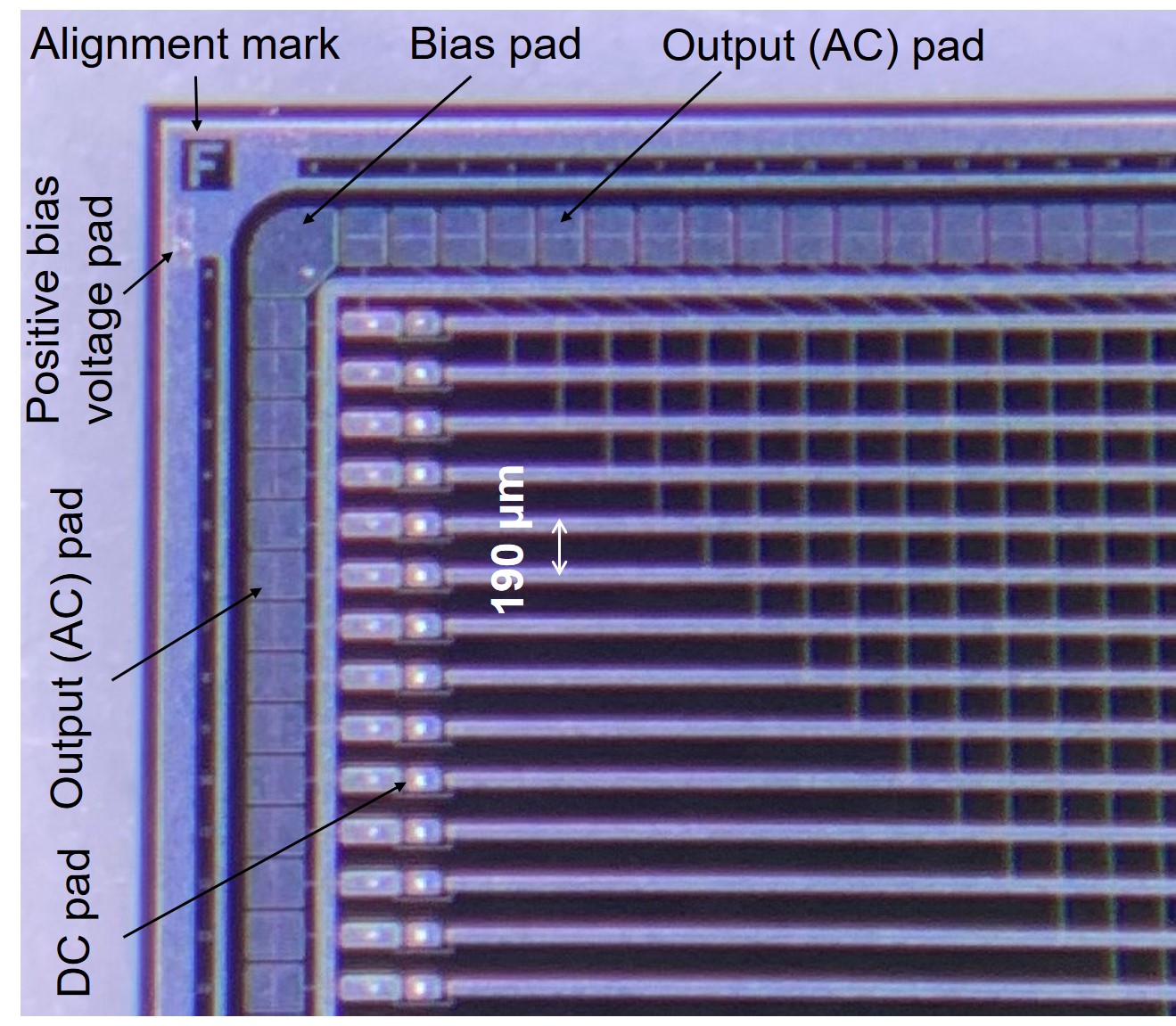}
    \caption{Silicon strip sensor used for the detector modules (left) and 
    the magnified image at the corner of the sensor (right).}
    \label{fig:sensor}
  \end{center}
\end{figure}

A silicon strip sensor developed for at the J-PARC muon $g-2$/EDM 
experiment is used in this module (figure~\ref{fig:sensor}).
The sensor is a single-sided p-on-n silicon
strip sensor with a double-metal structure made by Hamamatsu Photonics K. K.
and is now commercially available as a model S13804.
The size of the sensor is 98.77~mm $\times$ 98.77~mm with a thickness of 320~$\mu$m.
The active area is 97.28~mm $\times$ 97.28~mm. The sensor has two columns of
512 strips with a pitch of 190~$\mu$m and a length of 48.365~mm. 
The size of strips is determined to constrain the hit
rate less than 2~MHz and the estimated maximum hit rate at
the J-PARC muon $g-2$/EDM experiment with this strip size is 1.4~MHz per strip.
Strip sensor signals are read out from AC-coupled
output pads. Coupling capacitance between a p$^{+}$ strip 
and a read out strip is measured to be 170~pF. Because of double-metal
structure, signals can be read out from the
orthogonal direction as well. This
enables us to use the same design of the readout system for
horizontal or vertical strips since the sensor shape is square. By stacking two sensors with different
strip directions, two-dimensional coordinates of charged particles
passing through two sensors can be determined.
To construct 40 detection planes with two-dimentional coordinates for the J-PARC muon $g-2$/EDM experiment,
640 sensors are used in total.

The p$^{+}$ strips are connected to a common bias ring via
a polysilicon resister of the resistance of $10\pm 5$~M$\Omega$.
The bias ring is connected to ground via a bias pad on the surface.
The positive bias voltage is applied from either the back plane
of the sensor or pads on the surface.
The full depletion voltage was measured
to be from 65~V to 95~V depending on the sensor 
according to the measurements by the manufacturer
and sensors are usually operated with the bias voltage
above 120~V. The capacitance of one strip to the back plane
was measured to be about 17~pF above the full depletion voltage.
A typical total leakage current of the sensor is about 0.2~$\mu$A
with the bias voltage above the full depletion voltage at room temperature. 

\subsection{Front-end electronics}

\begin{figure}[htbp]
  \begin{center}
  \includegraphics[height=0.24\textheight]{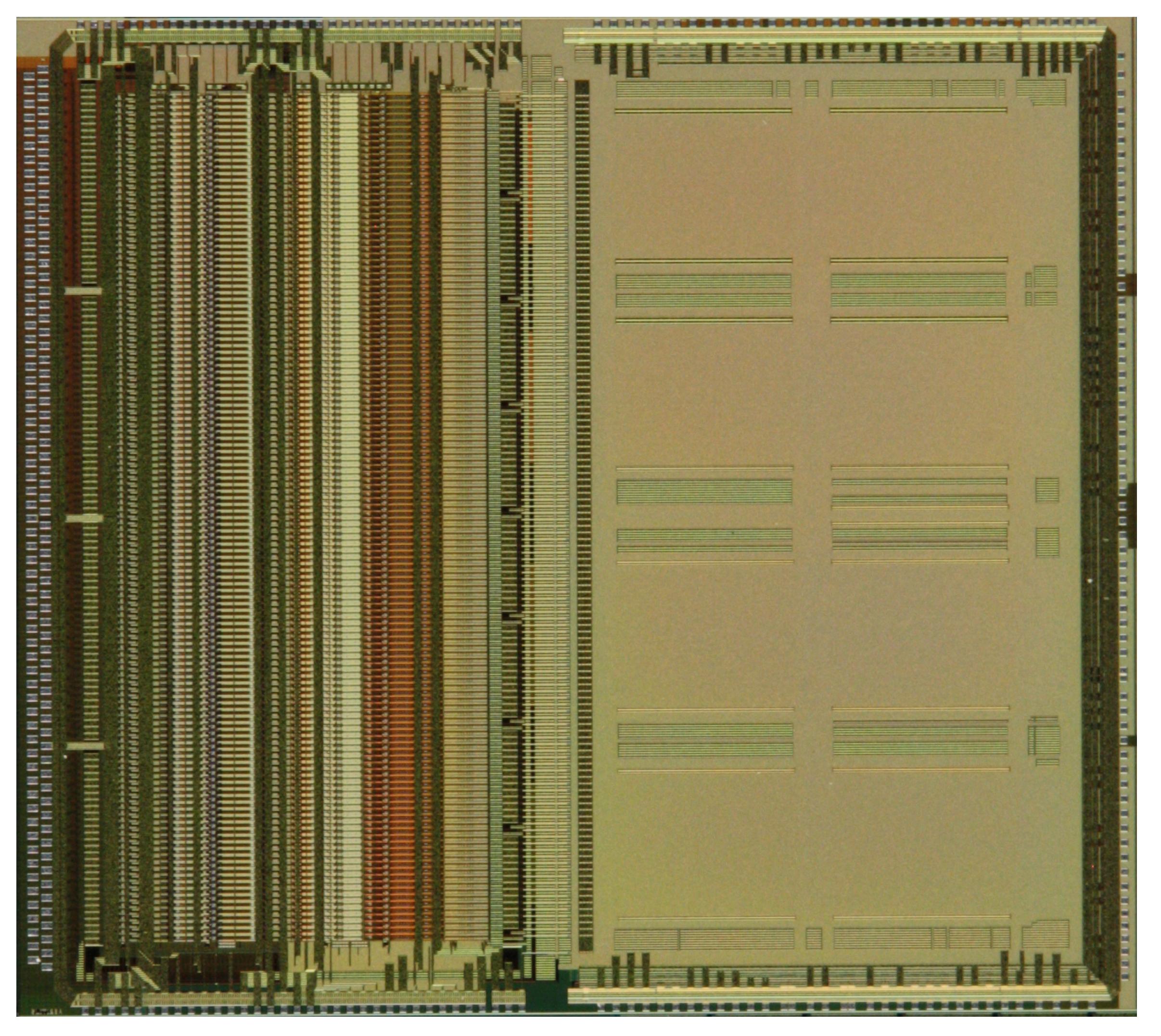}
  \caption{SliT128A chip.}
  \label{fig:SliT128A}
  \end{center}
\end{figure}

\begin{figure}[htbp]
  \begin{center}
  \includegraphics[width=0.95\textwidth]{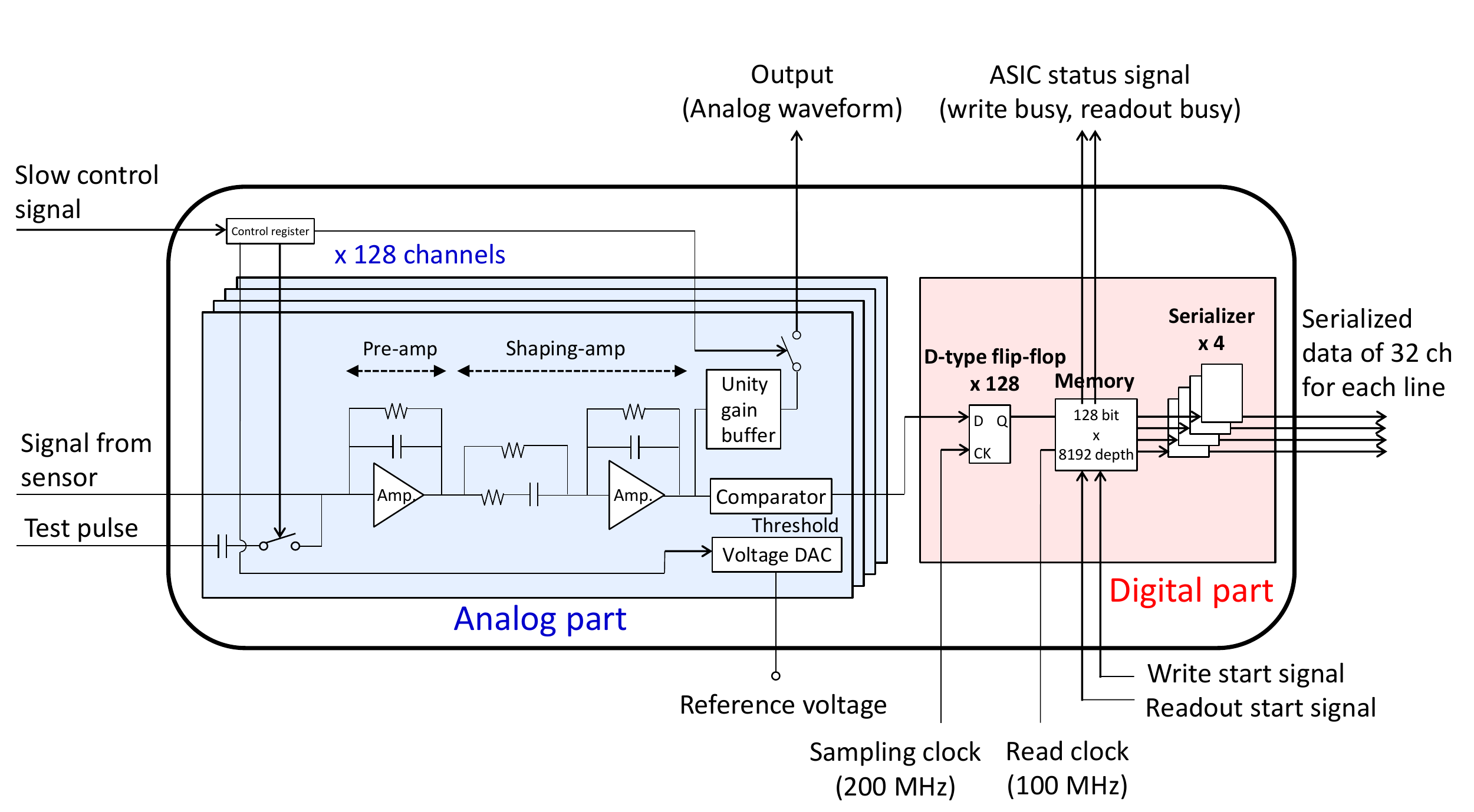}
  \caption{Schematic logic diagram of analog and digital circuits in a SliT128A chip.}
  \label{fig:SliT128A_logic}
  \end{center}
\end{figure}

The signals from the silicon strip sensors are read out by front-end boards 
with application specific integrated
circuits (ASICs) named ``SliT128A'',
which is a prototype of the front-end ASIC for the J-PARC 
muon $g-2$/EDM experiment.
The SliT128A is implemented in the Silterra 0.18-$\mu$m CMOS process and has a chip size of 9.0~mm~$\times$~10.0~mm with 128 readout channels
as shown in figure~\ref{fig:SliT128A}.
The schematic logic diagram of the SliT128A chips is shown in figure~\ref{fig:SliT128A_logic}.
Each channel has an analog part that consists of a preamplifier, a shaping amplifier, 
and a comparator with a reference voltage set by a 6-bit Digital-to-Analog Converter (DAC).
A charge of 3.84 fC~(1~MIP) is converted to a signal with a pulse width of about 180~ns by the preamplifier and the shaping amplifier.
The threshold value at the comparator is adjustable channel by channel using 
a 6-bit DAC to compensate for channel-to-channel threshold variations.
A monitor line is prepared to examine analog waveforms at the output of the shaping amplifier.
The control register enables the test pulse input line and the monitor line and adjusts the threshold DAC.

A digital part consists of 128 D-type flip-flops, a buffer memory and four serializers.
The binary signal from the comparator is sampled by the D-type flip-flop 
with 5~ns intervals provided by an external sampling clock with a frequency of 200~MHz.
When the SliT128 chip receives a write start signal, the chip starts to store the sampled data into the memory buffer.
For a pulsed (muon) beam, the write start signal is input to the SliT128A chip at the timing of 
pulsed (muon) beam 
coming based on a signal from an accelerator.
Since the memory buffer has a width of 128 bits and a depth of 8192 points,
we can store the data of 128 channels with a period of 40.96~$\mu$s ($=8192$~points~$\times$~5~ns).
A readout start signal is input after the end of the writing process is detected by the SliT128A status signal.
When the SliT128A chip receives readout start signal,
the data stored in the memory buffer is read out through serializers.
This readout must be finished before the next beam bunch coming.
To shorten read time,
four serializers are located
and each serializer serializes the data of 32 bits (or channels) times 8192 points
based on an external read out clock with a frequency of 100~MHz.

\begin{figure}[htbp]
  \begin{center}
  \includegraphics[width=0.9\textwidth]{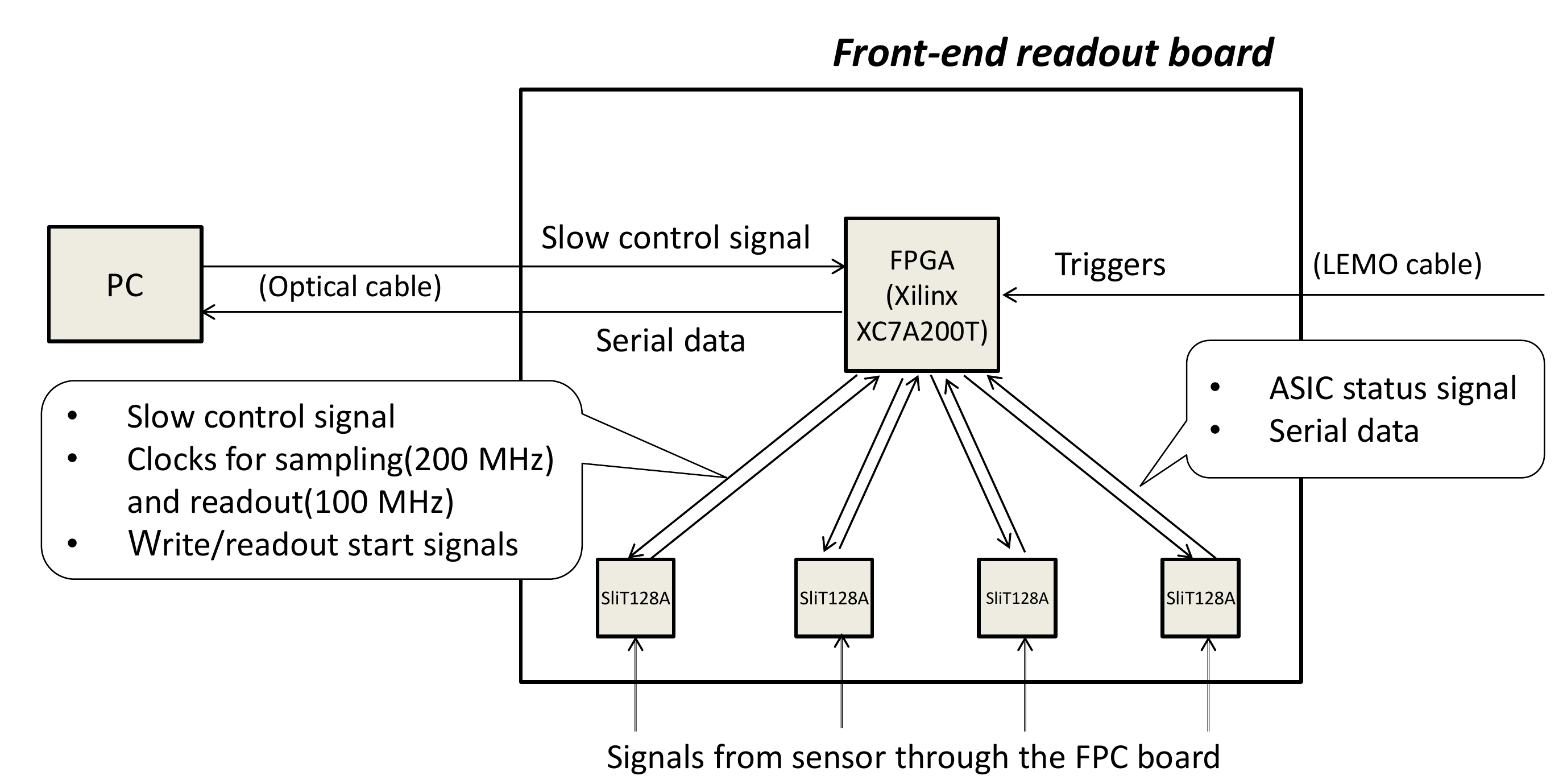}
  \caption{Configuration of the front-end readout board.}
  \label{fig:system_block}
  \end{center}
\end{figure}

The front-end readout board consists of four SliT128A chips,
a Xilinx Artix-7 XC7A200T field-programmable gate array (FPGA),
and a small form-factor pluggable (SFP) optical transceiver as shown in figure~\ref{fig:system_block}.
The data from the SliT128A chips are processed with the FPGA.
The start signals for the writing and readout processes can be input 
to the front-end readout board through the connector 
with the Nuclear Instrumentation Module (NIM) standard logic.
The data from four SliT128A chips are transferred to the first in, first out (FIFO) buffer
implemented inside the FPGA simultaneously in parallel,
when the start signal for the readout process is asserted.
The data of 32 channels with a depth of 8192 points comes per one readout line.
The data is saved into the FIFO with a format of 45 bits:
13 bits to represent hit timing and
32 bits to represent existence of hits for each channel.
The data can be compressed by the zero suppression algorithm:
if all 32 channels have no hits, the data is not stored to the FIFO  to suppress the amount of the data.
After completing the data transfer from the four SliT128A chips to the FIFO,
the data stored in the FIFO is serially transferred to a computer 
via an optical cable with Ethernet and SiTCP protocols~\cite{SiTCP}.

\subsection{Assembly}

A sensor is glued on a 1.6~mm-thick glass epoxy circuit board which has
 a hole in the active area of the sensor.
The positive bias voltage is applied from pads on the sensor surface, 
which is connected to the back plane of the sensor. The bias pads are
connected to ground.
To suppress parasitic capacitance, the bare SliT128A chips are directly placed 
on the front-end readout board and are wire bonded using 
25~$\mu$m-diameter aluminum (99\% Al with 1\% Si) wires. 
A sensor and SliT128A chips are connected by two flexible printed circuit (FPC) boards.
One FPC board connects 512 channels of a sensor to four SliT128A chips.
Wire bonding is used to connect a sensor with an FPC board and SliT128A chips 
with an FPC board.

\begin{figure}[htbp]
  \begin{center}
    \includegraphics[width=0.9\textwidth]{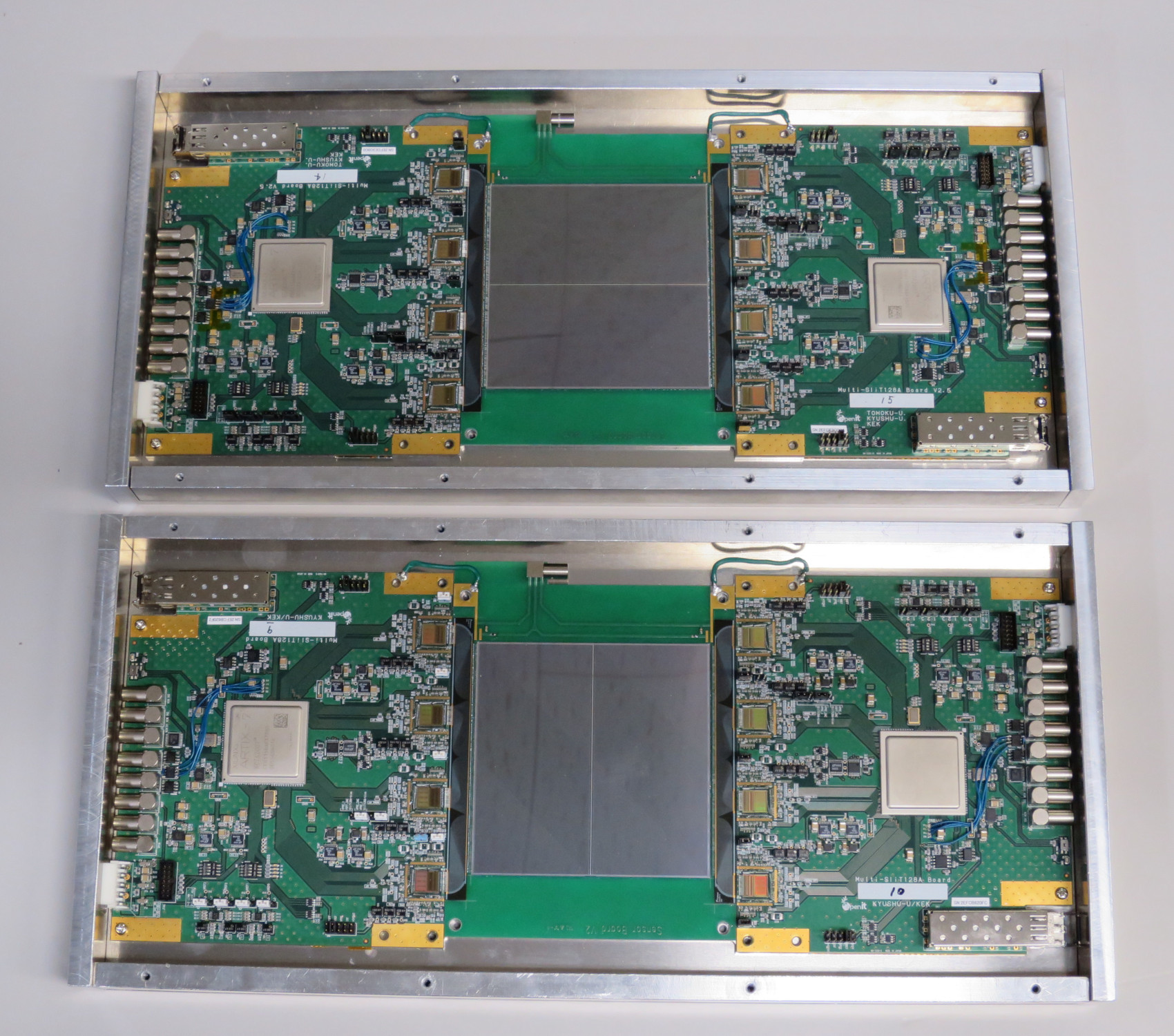}
    \caption{Detector modules with vertical (top) and with horizontal
      (bottom) strips.}
    \label{fig:modules}
  \end{center}
\end{figure}

A glass epoxy circuit board with a sensor and two readout boards are
glued on the same 3~mm-thick aluminum plate, which has a hole in the active area of the sensor.
Two types of modules were constructed;
one with strips in vertical 
and the other with strips in horizontal direction (figure~\ref{fig:modules}).
By stacking two types of modules successively, two-dimensional coordinates
of charged particles passing through two modules can be determined.

\subsection{Data acquisition system}

\begin{figure}[htbp]
  \begin{center}
  \includegraphics[width=0.8\textwidth]{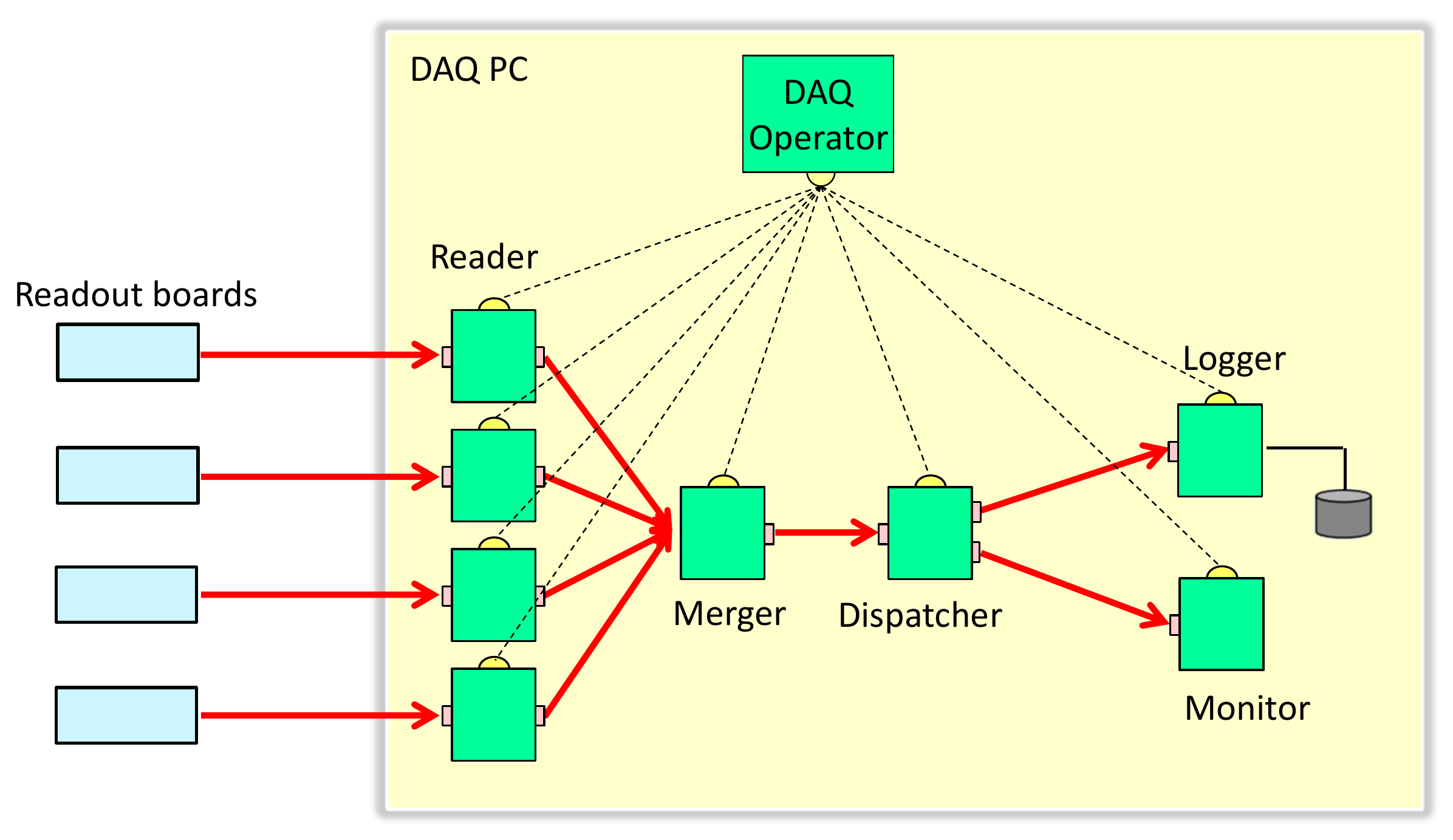}
  \caption{Structure of the DAQ system based on the DAQ-Middleware.}
  \label{fig:DAQ_system}
  \end{center}
\end{figure}

The data acquisition (DAQ) system is constructed based on 
DAQ-Middleware~\cite{DAQMiddleware,DAQMiddleware2},
which is a software framework for network distributed DAQ software.
The framework consists of DAQ-Components and a DAQ-Operator.
The DAQ-Component is a base unit that can run fully independently of each other.
Users can develop a data acquisition path by connecting DAQ-Components.
The DAQ-Operator controls DAQ-Components by sending the signals to 
each DAQ-Component, such as a start and a stop.
Figure~\ref{fig:DAQ_system} shows a structure of the DAQ system.
There are five types of the DAQ-Components in the DAQ system for this module: reader, merger, dispatcher, logger, and monitor.
Data from one front-end readout board is received by a reader component.
The data acquisition paths for each front-end readout board are merged by the merger component
and the merged path is split into two by the dispatcher component.
One path is connected with the storage component and the data is saved in the hard disk drive.
The other path is connected with the monitor component and
a part of the data is analyzed to monitor the status of the module and the DAQ system itself.
The maximum data rate from a front-end readout board is about 20 MB/s without the zero suppression algorithm.
It takes about 2.6~ms for the data transfer from four SliT128A chips 
to the FIFOs in the FPGA
and it takes about 5.3~ms per front-end readout board without the zero-suppression algorithm 
for the data transfer from the FIFOs in the FPGA to a personal computer.
One detector module can be operated up to a repetition rate of 75 Hz without the zero-suppression algorithm
and it satisfies the requirement of the muon beam rate at the J-PARC.

\section{Quality control}
\begin{figure}[htbp]
  \begin{center}
    \includegraphics[width=0.49\textwidth]{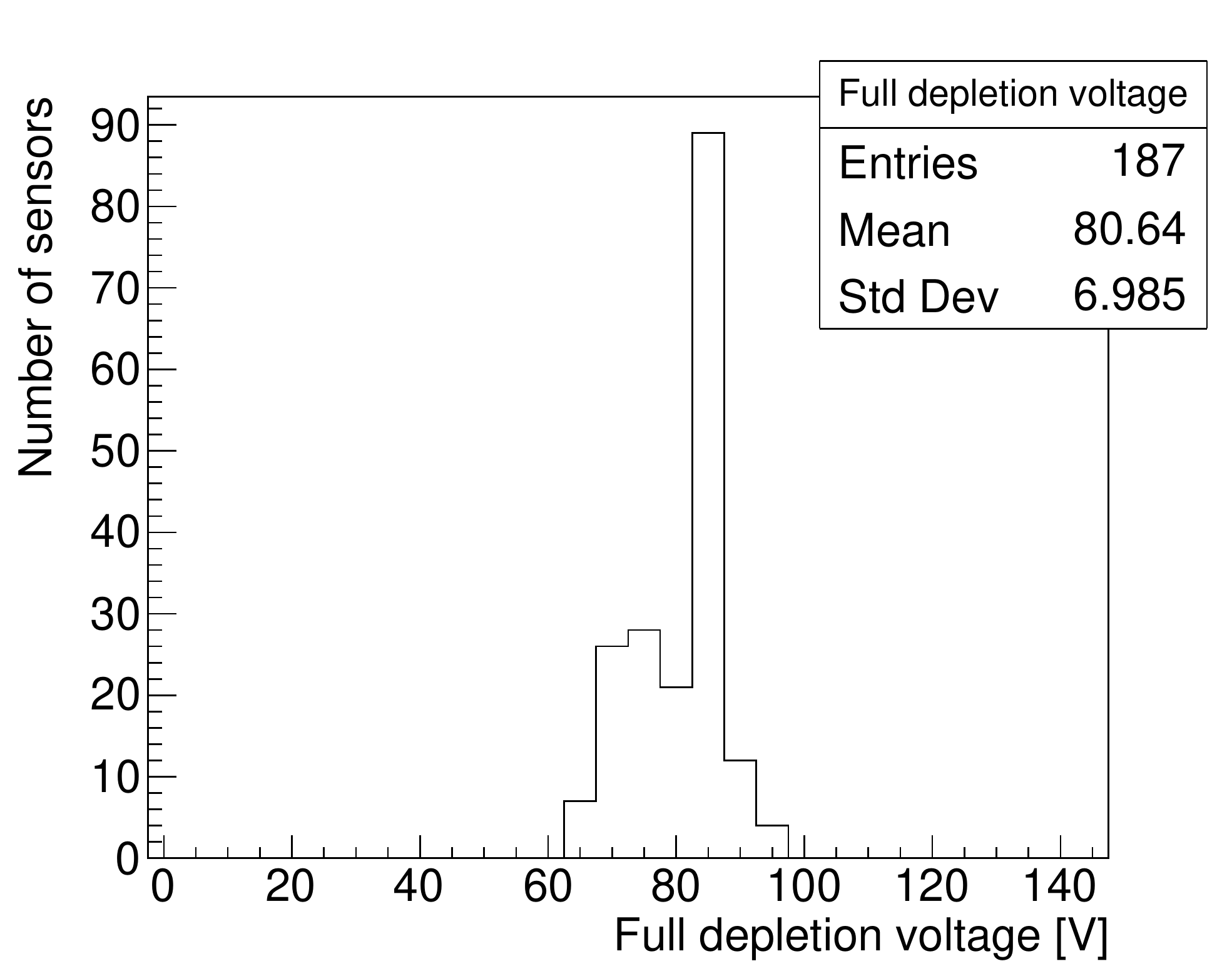}
    \includegraphics[width=0.49\textwidth]{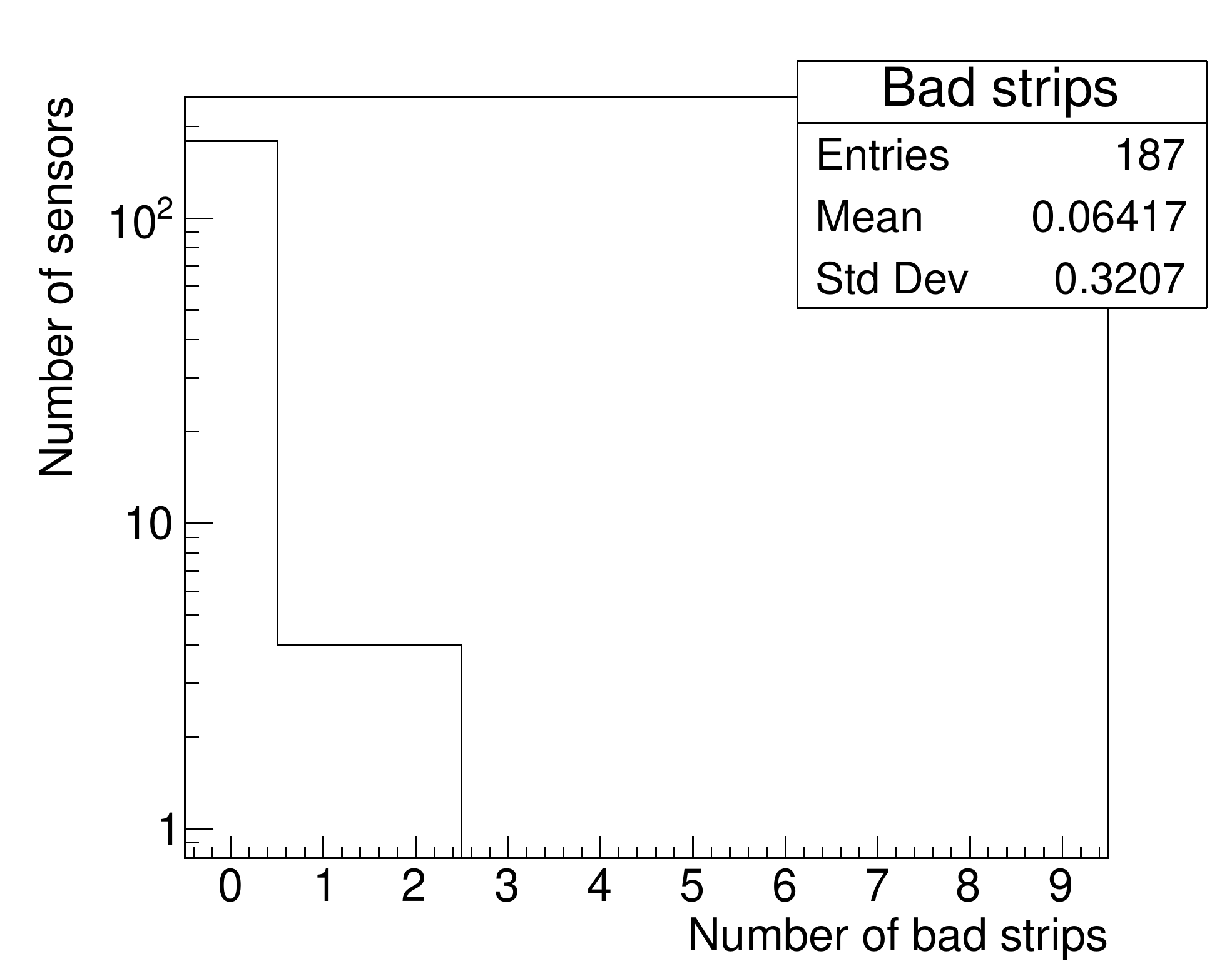}
    \caption{The full depletion voltage (left) and the statistics of bad strips
      per sensor (right) for 187 delivered sensors measured by 
      Hamamatsu Photonics K. K.}
    \label{fig:FDV_BadStrip}
  \end{center}
\end{figure}

\begin{figure}[htbp]
  \begin{center}
    \includegraphics[width=0.49\textwidth]{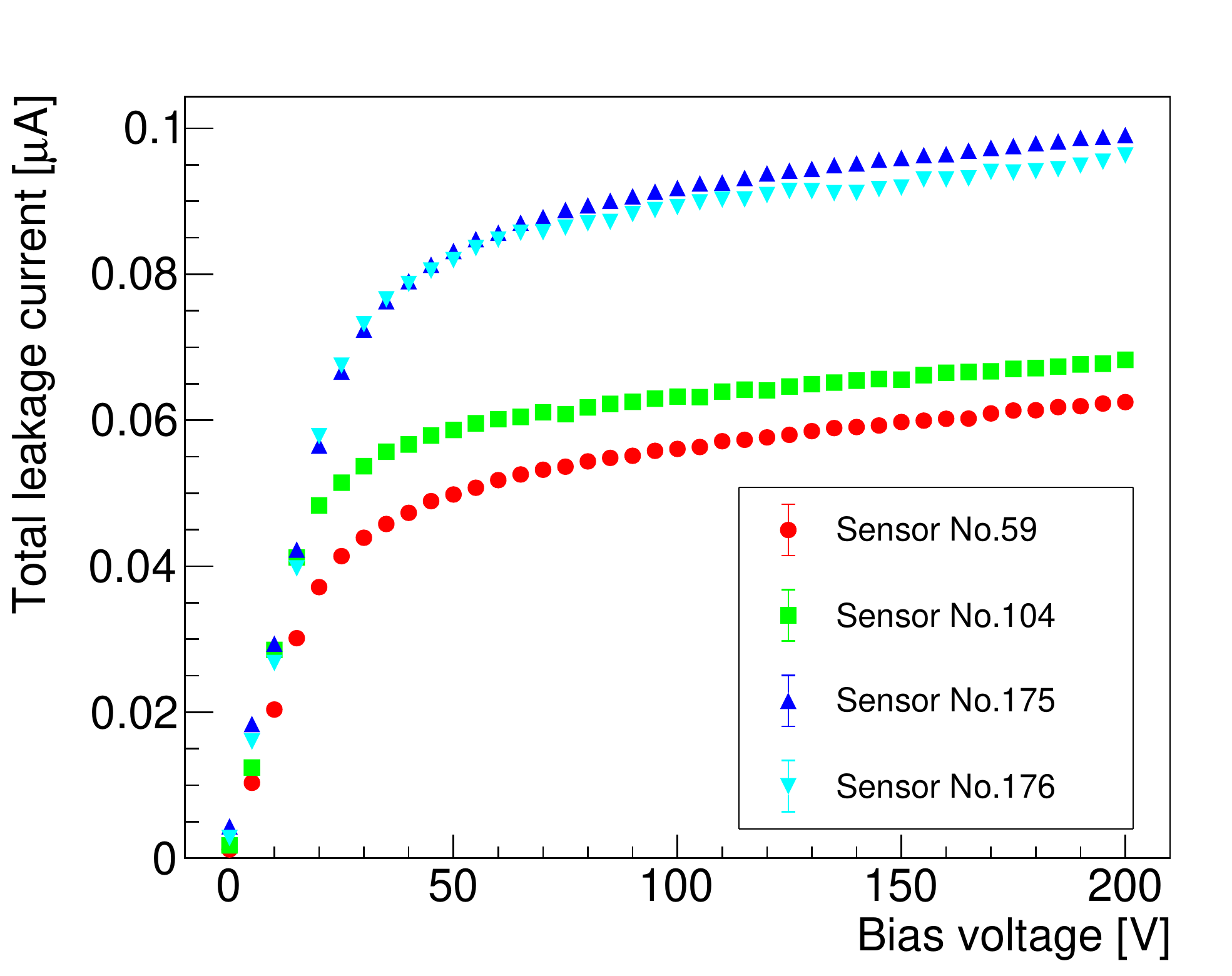}
    \includegraphics[width=0.49\textwidth]{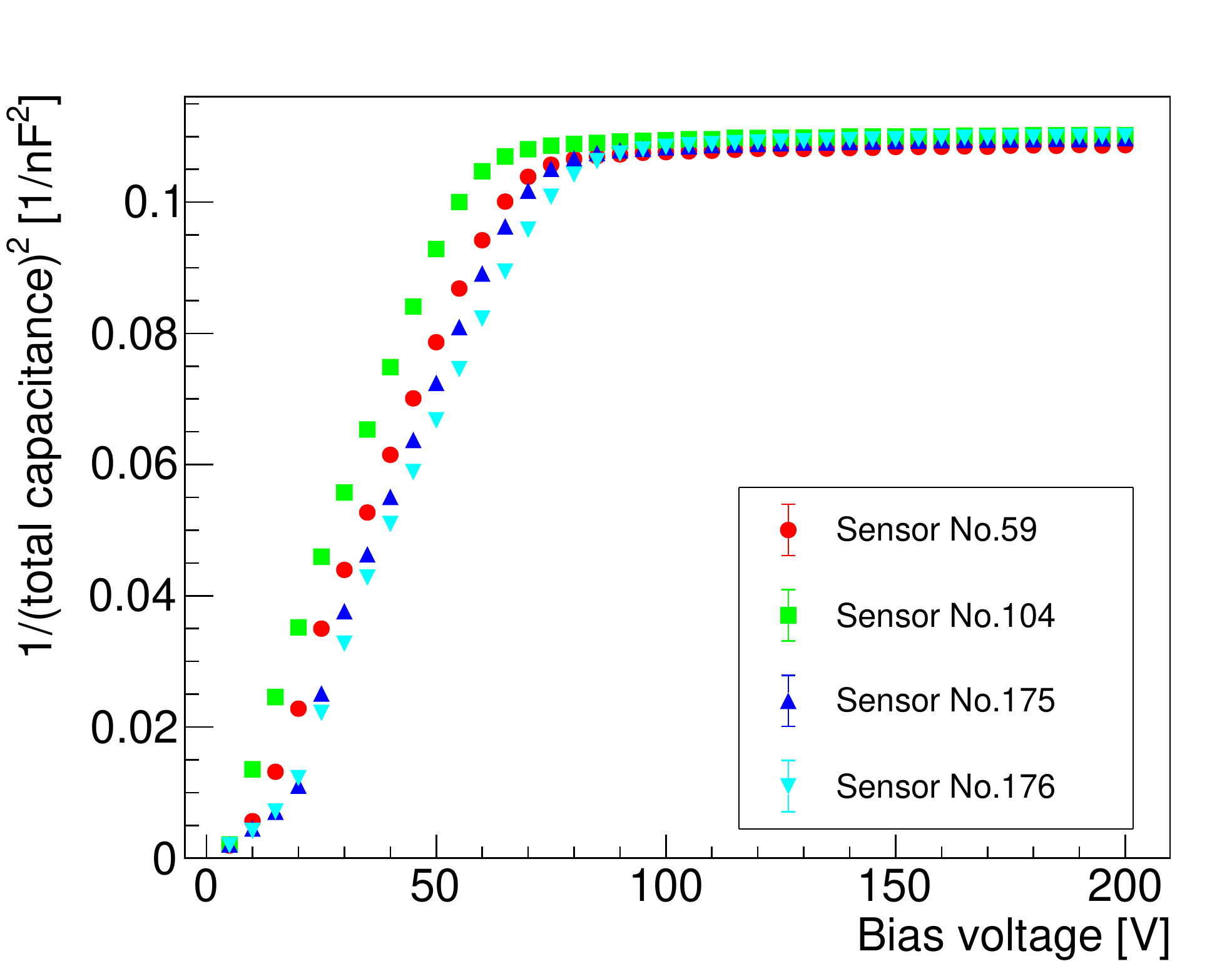}
    \caption{The total leakage current (left) and capacitance (right) 
      as a function of the bias voltage of sensors, which were used in 
      detector module assembly. The total leakage current 
      value is corrected to the measurement at 20~$^{\circ}$C.
    }
    \label{fig:IV_CV_measurement}
  \end{center}
\end{figure}

Basic performance of silicon strip sensors 
was measured by the manufacturer before shipment.
The total leakage current and capacitance of the sensor
 were measured as a function of the
bias voltage. Any strip defects (e.g., a shortage of 
AC-DC strips, or an isolation of strips)
were recorded and only the sensors whose defective
strip ratios were less than 5\% were delivered.
The full depletion voltage and the statistics of bad strips
for 187 sensors
measured by the manufacturer are shown in figure~\ref{fig:FDV_BadStrip}.
The total leakage current and capacitance were also 
measured in our laboratory (figure~\ref{fig:IV_CV_measurement})
and they were consistent with the measurements by the manufacturer.
The total leakage current 
was also measured at each step of module assembly 
to find any damage that appeared during assembly.

\begin{figure}[htbp]
  \begin{center}
    \includegraphics[width=0.5\textwidth]{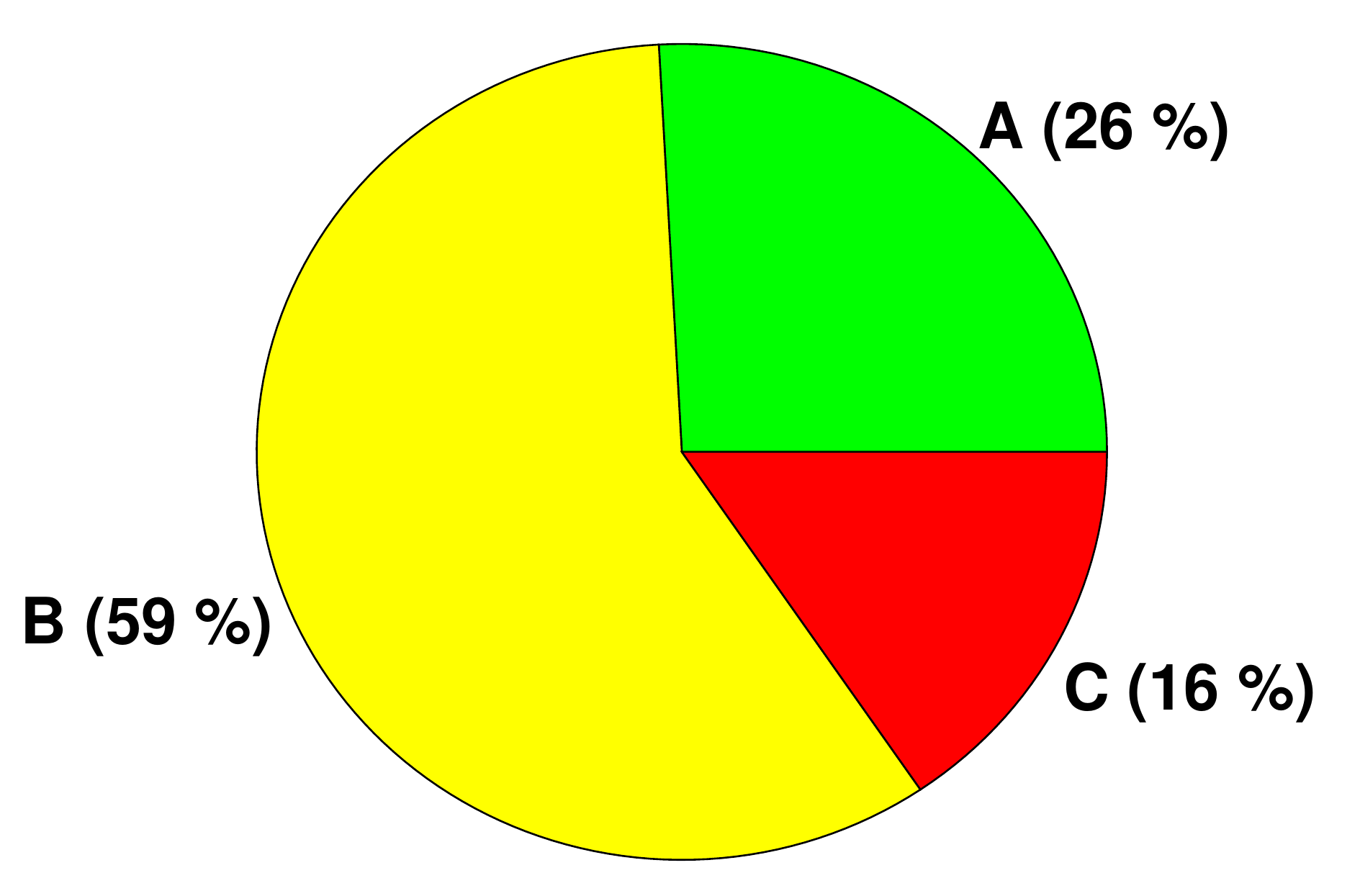}
    \caption{The ratio of quality of SliT128A measured for 115 chips. }
    \label{fig:ASIC_quality}
  \end{center}
\end{figure}

Basic performance of SliT128A chips was measured
before assembly to the readout board. A probe card was used to deliver signals
from a SliT128A chip and power supplies to a SliT128A chip. 
Quality of SliT128A chips were categorized into a rank of A, B and C.
Chips without any defects are categorized as a rank A.
Chips with a noise level greater than 1250 electrons on average 
or more than five noisy (comparator output is always high) or dead (no comparator output) channels 
are categorized as a rank B.
Chips with severe defects such as malfunction of comparator output in more than 32 channels
are
categorized as a rank C.
The ratio of quality of SliT128A measured for 115 chips
is shown in figure~\ref{fig:ASIC_quality}. SliT128A chips without
any defects (rank A) were used for
module assembly. 
The same tests were performed after connecting
SliT128A chips to a readout board and after connecting
a sensor to SliT128A chips. The readout boards without any defective
SliT128A chips were used for module assembly.

The pull test was performed on wires bonded on
spare parts of detector modules prior to practical assembly.
The bonding parameters were optimized to satisfy the
pull strength greater than 60~mN at any bonding positions 
(typically, greater than 80~mN).
The same bonding parameters were used at practical assembly.

\section{Detector performance}
\label{sec:detector_peformance}

\subsection{Laboratory measurements}

\begin{figure}[htbp]
  \begin{center}
    \includegraphics[width=0.6\textwidth]{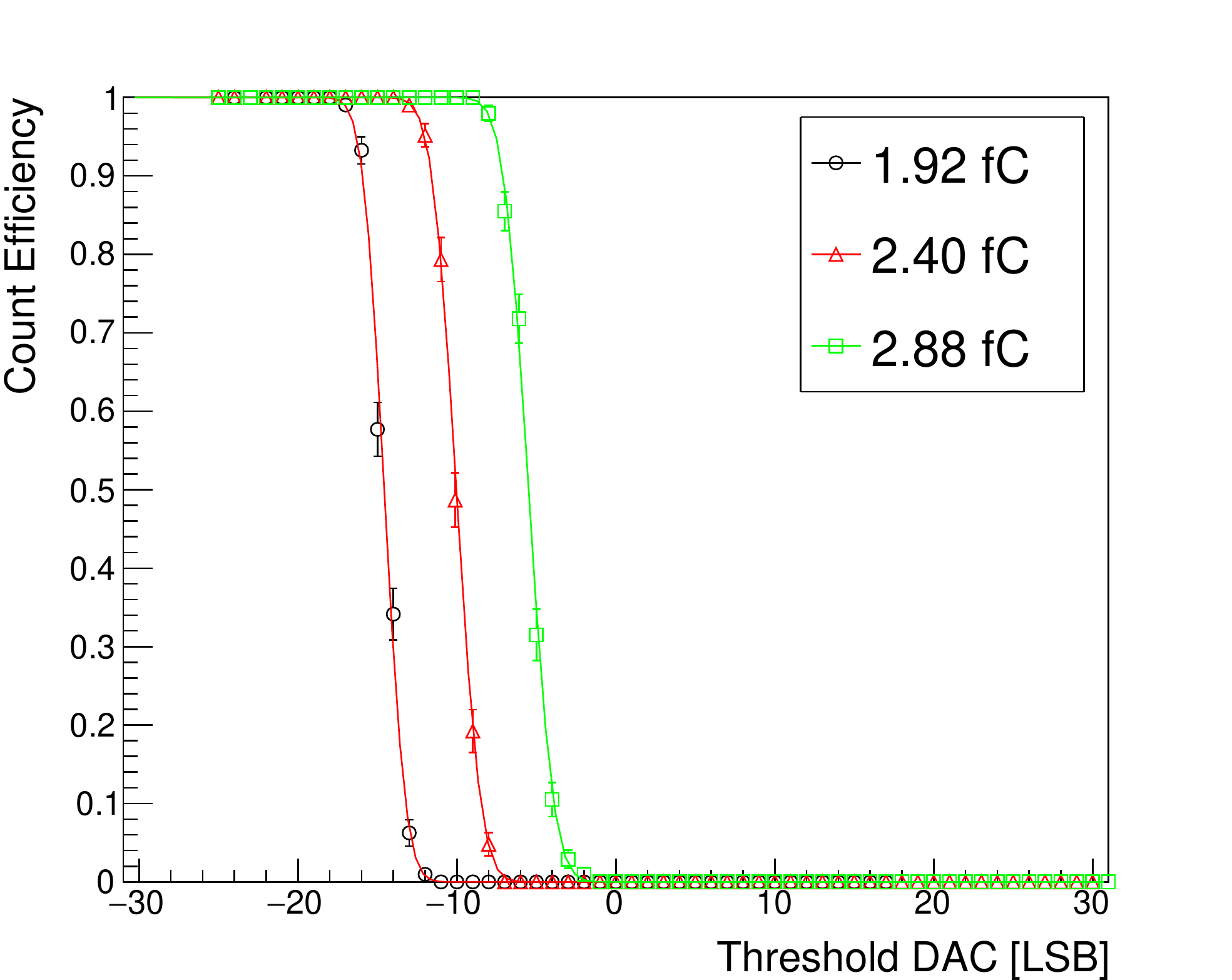}
    \caption{Examples of the counting efficiency of one readout channel as a function of the threshold with injected charges of 1.92~fC (open black circles), 2.40~fC (open red triangles), and 2.88~fC (open green squares), respectively.
    The threshold is shown in the unit of the least significant bit (LSB) of the DAC.}
    \label{fig:scurve}
    \includegraphics[width=0.6\textwidth]{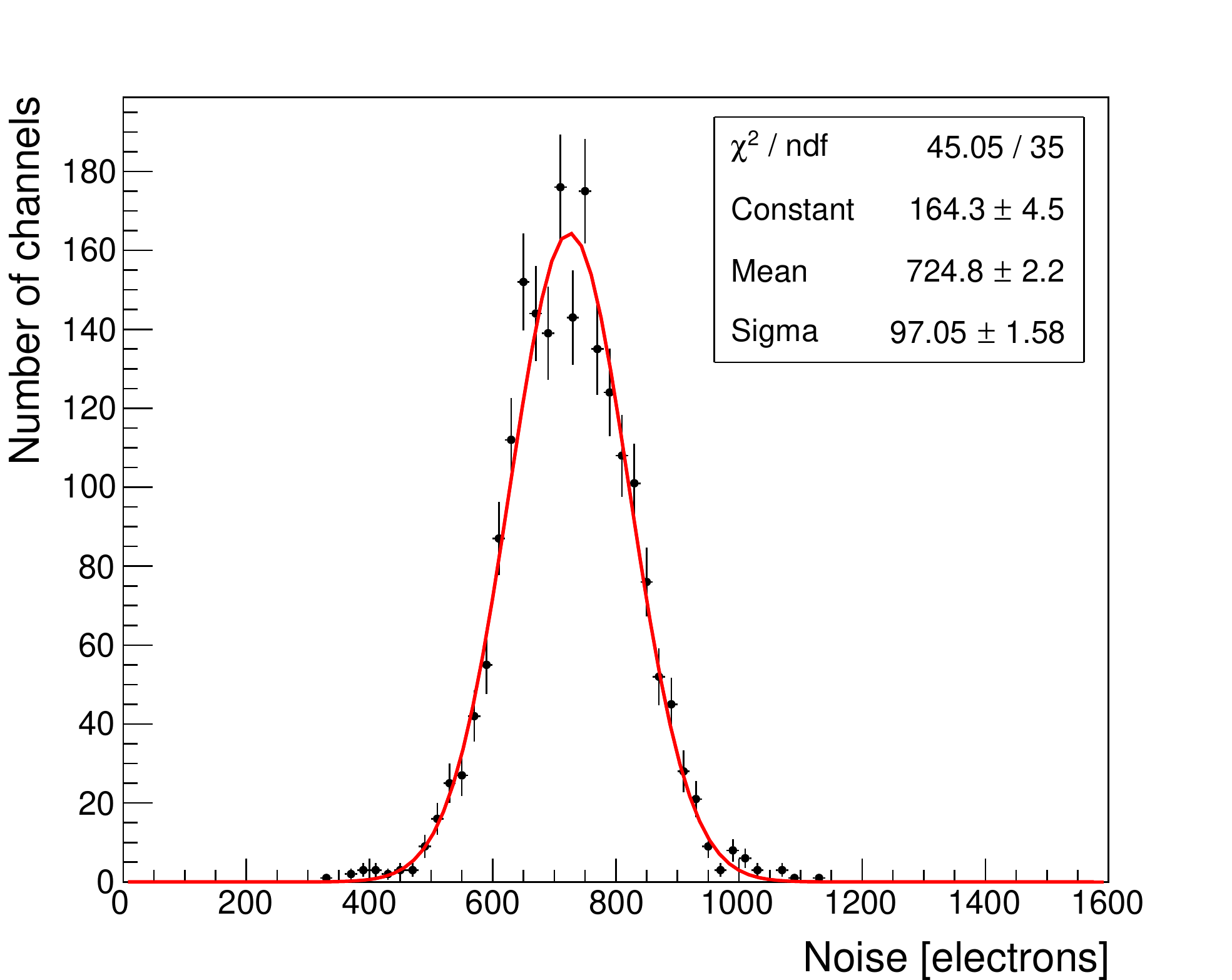}
    \caption{Equivalent noise charge distribution of all readout channels in 
            two detector modules. A Gaussian function is fitted to data.}
    \label{fig:enc}
  \end{center}
\end{figure}

Before the test with a positron beam, 
the noise, the time over threshold (ToT), and the time walk of the detector module
were evaluated
using test pulses generated by a function generator.
The noise can be evaluated from the threshold scan with different 
injected test pulse charges.
A complementary error function is fitted to the counting efficiency as a function of the threshold voltage.
The signal amplitude corresponds to the threshold at 50\% efficiency and the noise level
corresponds to the difference of thresholds at efficiency of 50\% and 84.3\%.
Figure~\ref{fig:scurve} shows examples of the counting efficiency curves of one readout channel.
Figure~\ref{fig:enc} is the equivalent noise distribution of all readout channels in two detector modules.
The mean of the equivalent noise charge (ENC) is measured to be 725~electrons with a detector capacitance of the silicon strip sensor of 17~pF. 
For the actual detector to be used at the J-PARC muon $g-2$/EDM experiment,
the detector capacitance is estimated to be about 30~pF because of
a floating capacitance of long signal transport line.
The dependence of the noise on the detector
capacitance is evaluated to be 33~electrons/pF according to the circuit simulation.
The mean of the ENC is estimated to be 1150~electrons at the detector capacitance of 30~pF.

\begin{figure}[htbp]
  \begin{center}
    \includegraphics[width=0.49\textwidth]{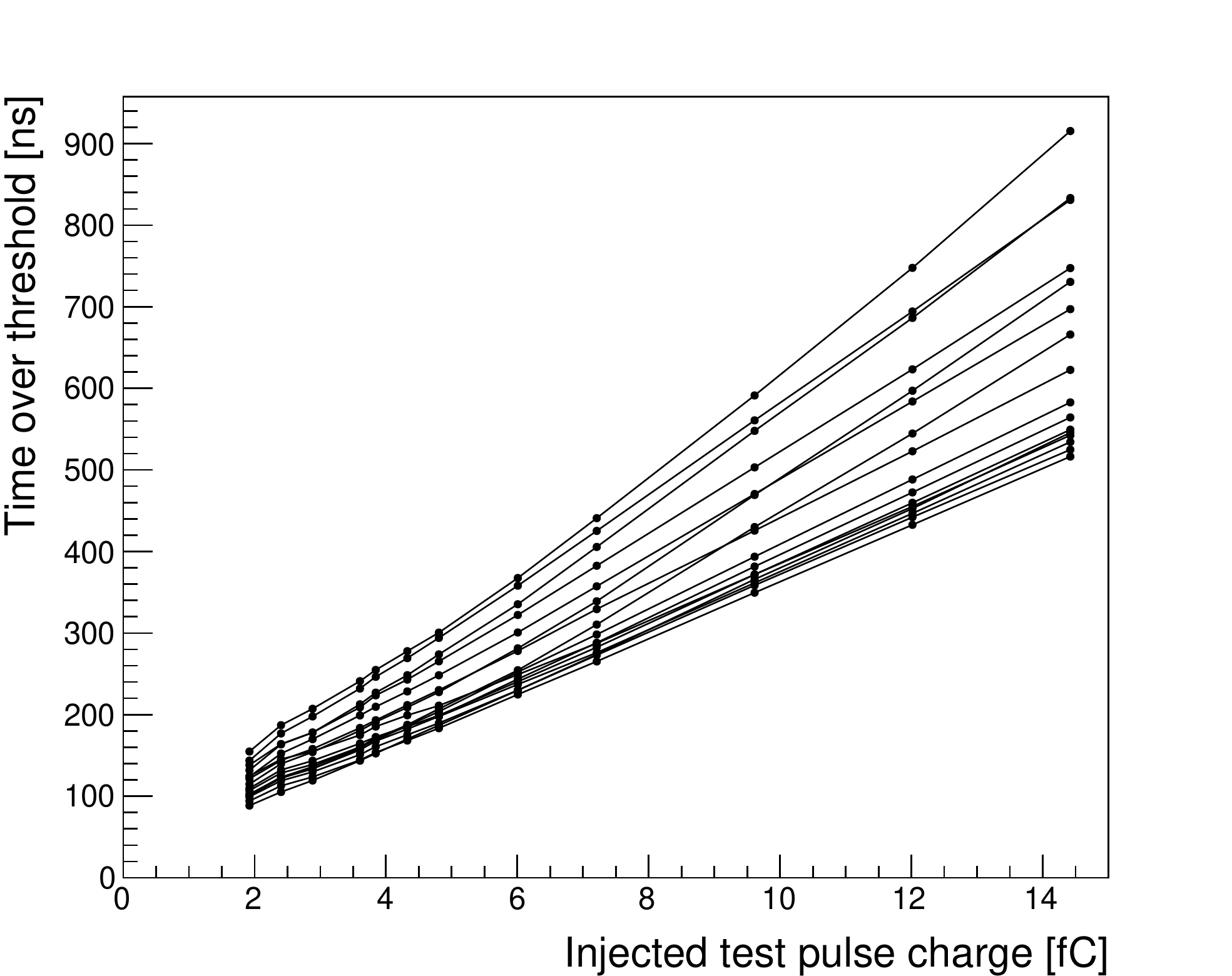}
    \includegraphics[width=0.49\textwidth]{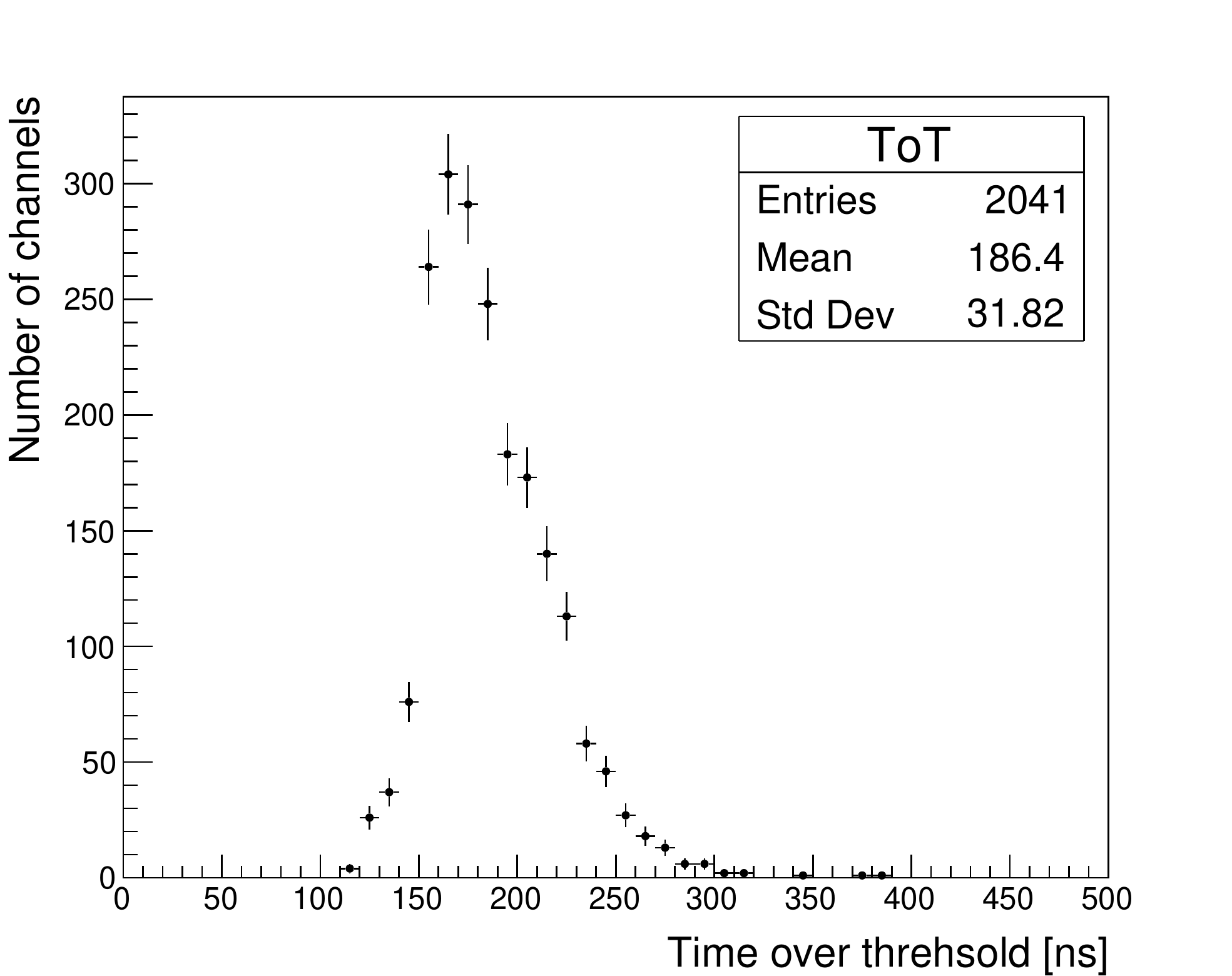}
    \caption{
ToT as a function of the injected test pulse charge of typical 16 readout channels (left)
and the ToT distribution with an injected test pulse charge of 3.84~fC for all readout channels in two detector modules (right).
    }
    \label{fig:ToT}
  \end{center}
\end{figure}

Based on the result of the threshold scan, the threshold voltage is set to match 1.15~fC which corresponds to
30\% of MIP
charge (3.84~fC) level in each channel.
The ToT, which is the width of the comparator output, is measured as a function of the test pulse charge 
at this threshold.
Figure~\ref{fig:ToT} shows the ToT distribution with an injected test pulse charge of 3.84~fC.
The mean value of the ToT is 186~ns. According to this measurement result,
detectors are estimated to tolerate up to a hit rate of $\sim$5~MHz on average
for a MIP charge signal.

\begin{figure}[htbp]
  \begin{center}
     \includegraphics[width=0.49\textwidth]{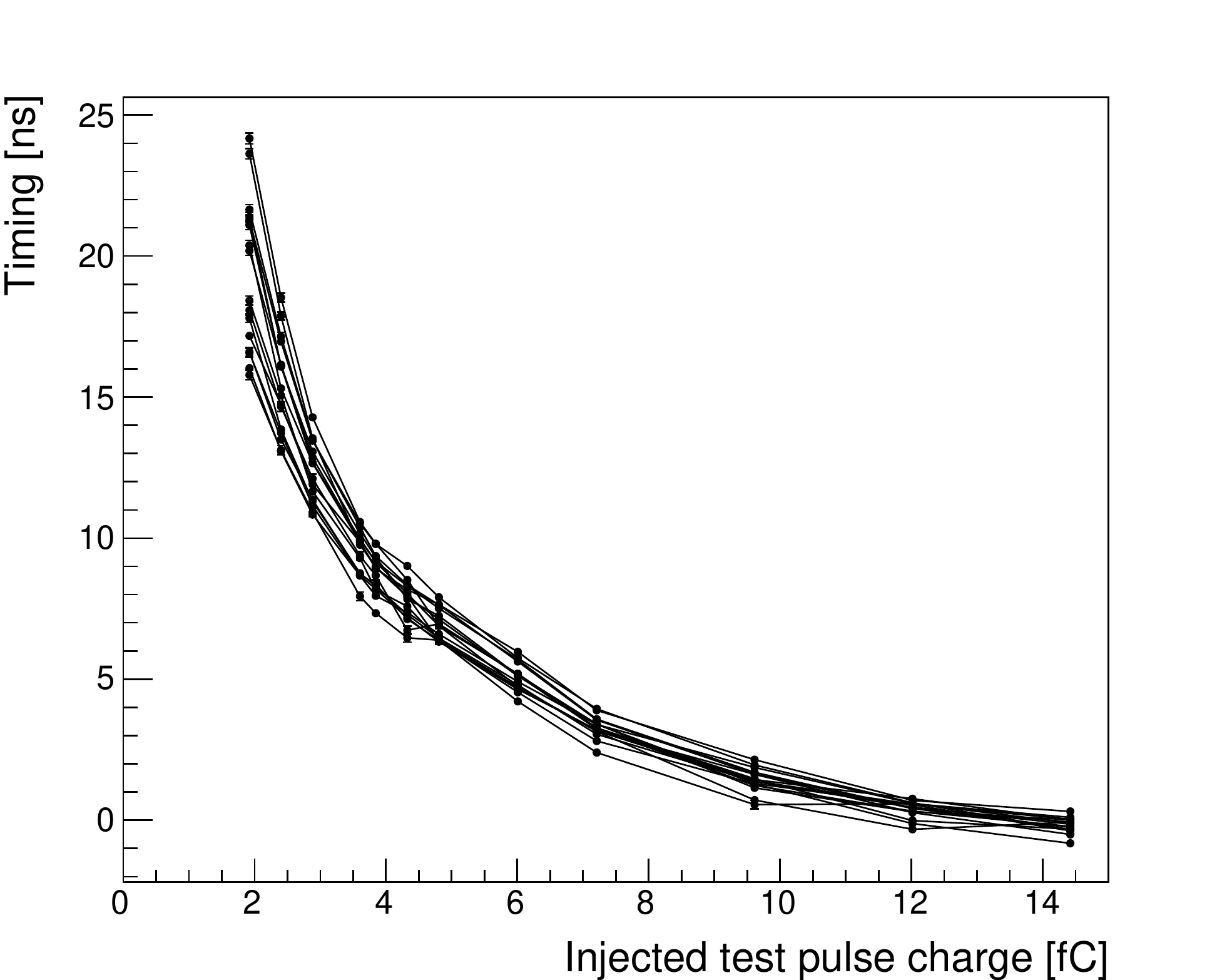} 
     \includegraphics[width=0.49\textwidth]{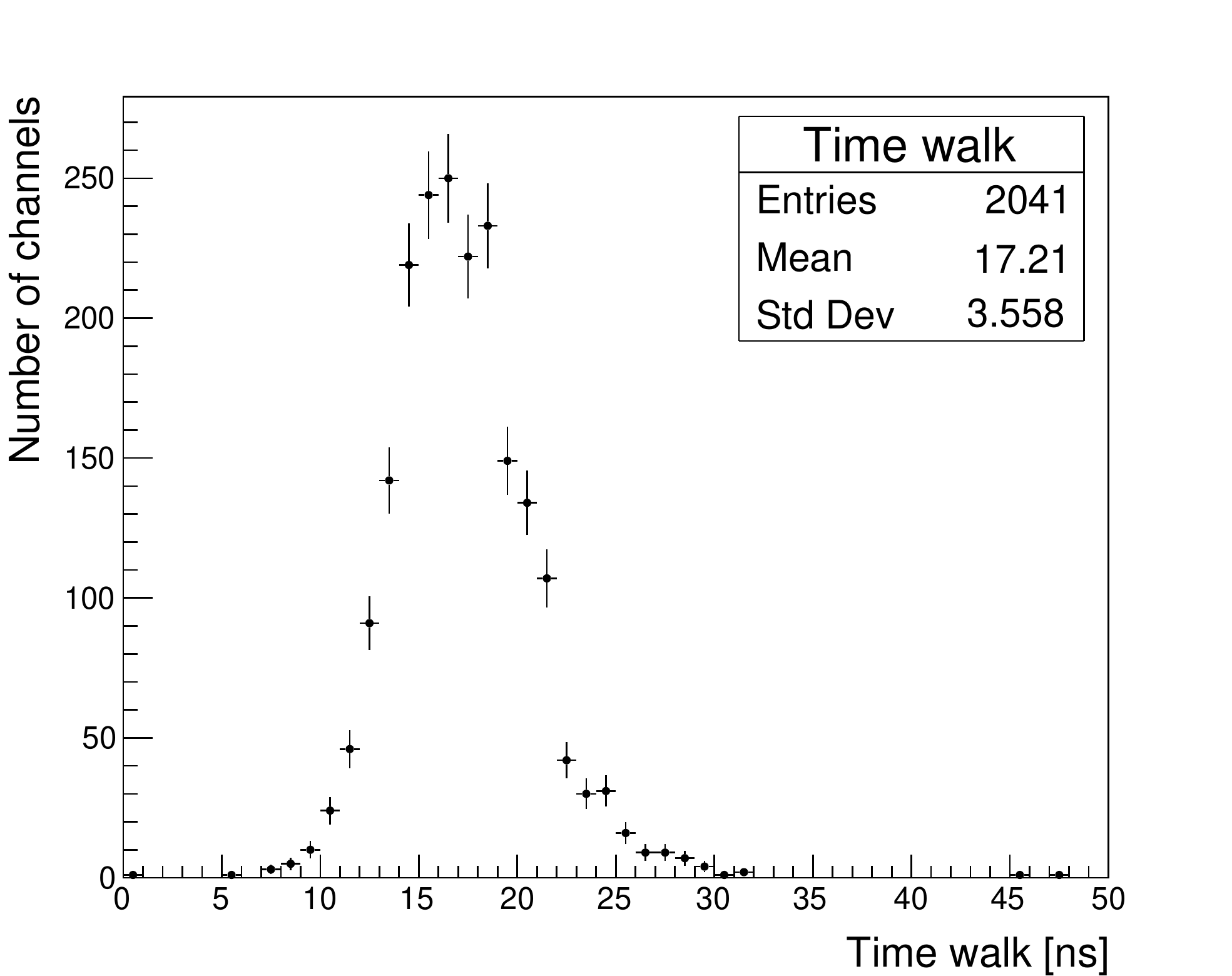} 
    \caption{
Leading edge timing of the comparator as a function of the injected test pulse charge of typical 16 readout
        channels (left)
and the distribution of the time walk over a signal range of 1.92-11.52~fC 
for all readout channels in two detector modules (right).}
    \label{fig:TimeWalk}
  \end{center}
\end{figure}

The time walk is defined as the maximum time variation in the comparator output over a signal range of 1.92-11.52~fC,
with the comparator threshold set to 1.15~fC.
Figure~\ref{fig:TimeWalk} shows the distribution of the time walk of two detector modules.
The mean value of the time walk is 17.2~ns. This value is larger than the requirement
at the J-PARC muon $g-2$/EDM experiment of 1~ns. This will be solved by the next
version of the front-end ASIC.
Time variation due to the time walk is corrected later to improve the time resolution.

\subsection{Beam test setup}

\begin{figure}[htbp]
  \begin{center}
    \includegraphics[width=0.95\textwidth]{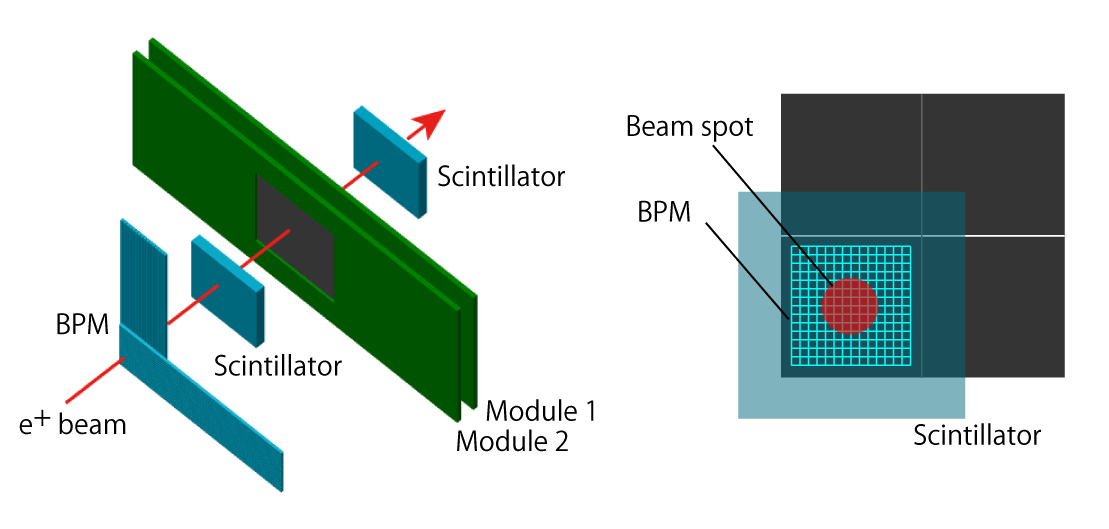}
    \caption{Perspective view of the detector setup in the beam test (left) and the view from the beam axis (right).}
     \label{fig:beamtestsetup}
  \end{center}
\end{figure}

The performance of the detector modules was evaluated
using a positron beam at the ELPH, Tohoku University~\cite{Ishikawa}.
The momentum-analyzed positron beam with the momentum of 730~MeV/$c$ was produced 
from 1.3~GeV electrons by a series of reactions of Bremsstrahlung and pair production.
The setup of the beam test is shown in figure~\ref{fig:beamtestsetup}.
Two detector modules whose strip directions are orthogonal were put perpendicular to the beam axis.
The module with horizontal strips (Module 1) is downstream of the beam line
and the module with vertical strips (Module 2)  upstream. 
Events were triggered by the coincidence of two scintillation
detectors sandwiching the silicon strip detector modules.
The size of the scintillator is 800~mm $\times$ 800~mm $\times$ 5~mm in width, height and thickness, respectively.
The coincidence rate was about 1~kHz.
For the coincidence trigger, 
the FPGAs on detector modules continuously send the start signals for the writing process 
at intervals of 50~$\mu$s, which is longer than buffer length of 40.96~$\mu$s.
This is to avoid injecting the start signal for the writing in the
state of writing process.
If the FPGAs detect the coincidence trigger signal, 
they stop the continuous start signal for the writing process and send the start signal for the readout process.
To reduce the trigger rate lower than the acceptable rate of the DAQ system, the trigger rate is prescaled to about 90~Hz.
Almost only one positron hit the detector modules in one event in this condition. 
A beam profiling monitor (BPM) consisting of two layers of 14 scintillating fibers was 
installed to determine the incident position of positrons.
The fiber is made from 3~mm square cross-section scintillator, and the incident position is determined with a 3~mm resolution.
The center-of-the-beam position is set at the lower left of the modules as shown in figure~\ref{fig:beamtestsetup} (right).

\subsection{Measurement results}

\begin{figure}[htbp]
  \begin{center}
    \includegraphics[width=0.8\textwidth]{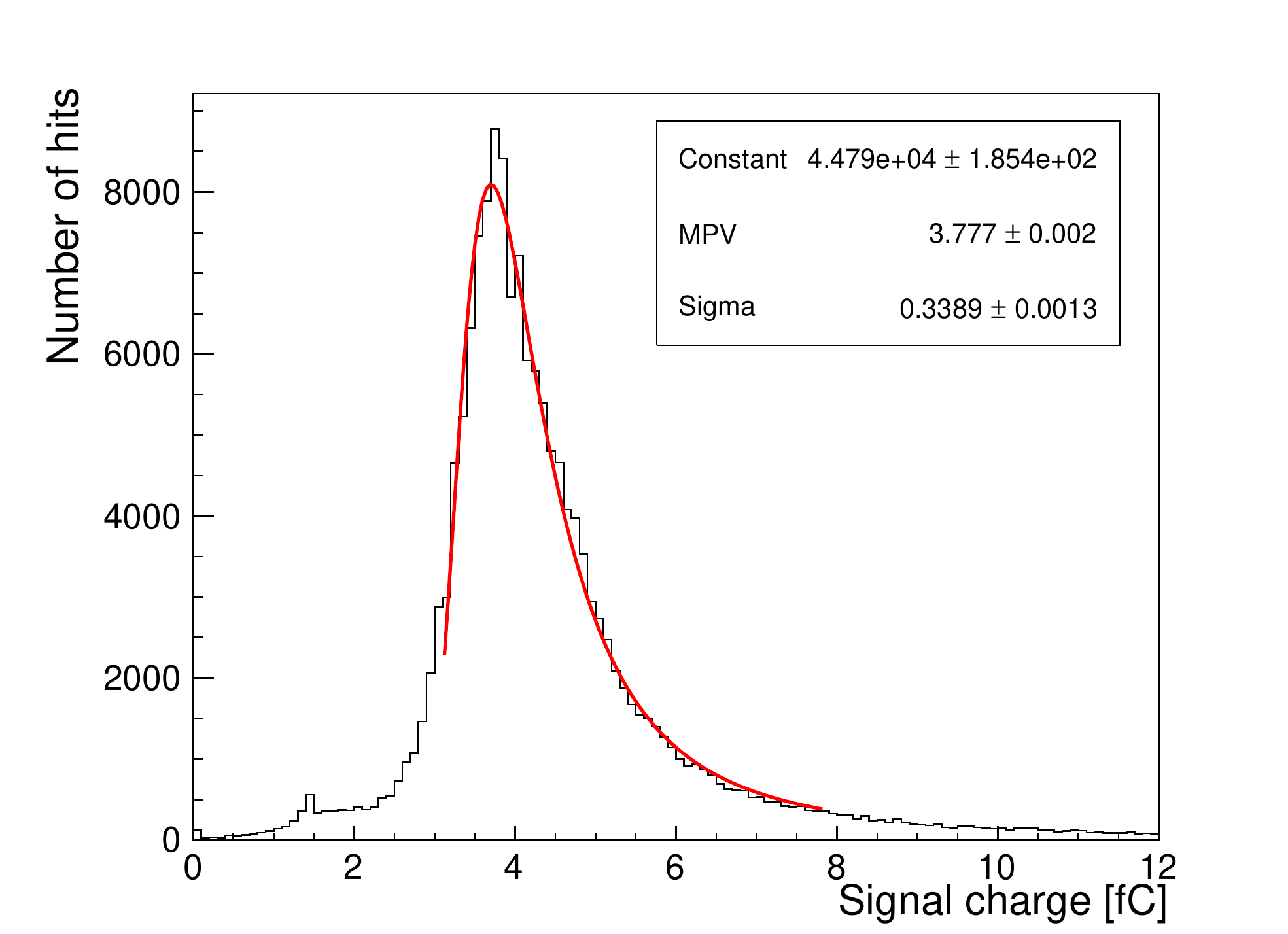}
    \caption{Charge distribution of the positron beam measured by two detector modules. 
    A Landau distribution is fitted to data.}
    \label{fig:charge_dist}
  \end{center}
\end{figure}

Figure~\ref{fig:charge_dist} shows the charge distribution of the beam signal measured
by two silicon strip detector modules.
The most probable value (MPV) of the charge distribution is $3.777 \pm 0.002$~fC, which
is consistent with the simulated MIP charge of 3.84~fC within an uncertainty on the charge
calibration (about 2\%).
The signal-to-noise ratio is derived to be $32.6$ at the detector capacitance
of 17~pF, by taking the ratio of 
the MPV of the charge distribution measured by the positron beam to the 
mean of the noise level measured by test pulses.
At the detector capacitance of 30~pF, the signal-to-noise ratio is estimated
to be 20.8.

\begin{figure}[htbp]
  \begin{center}
    \includegraphics[width=0.75\textwidth]{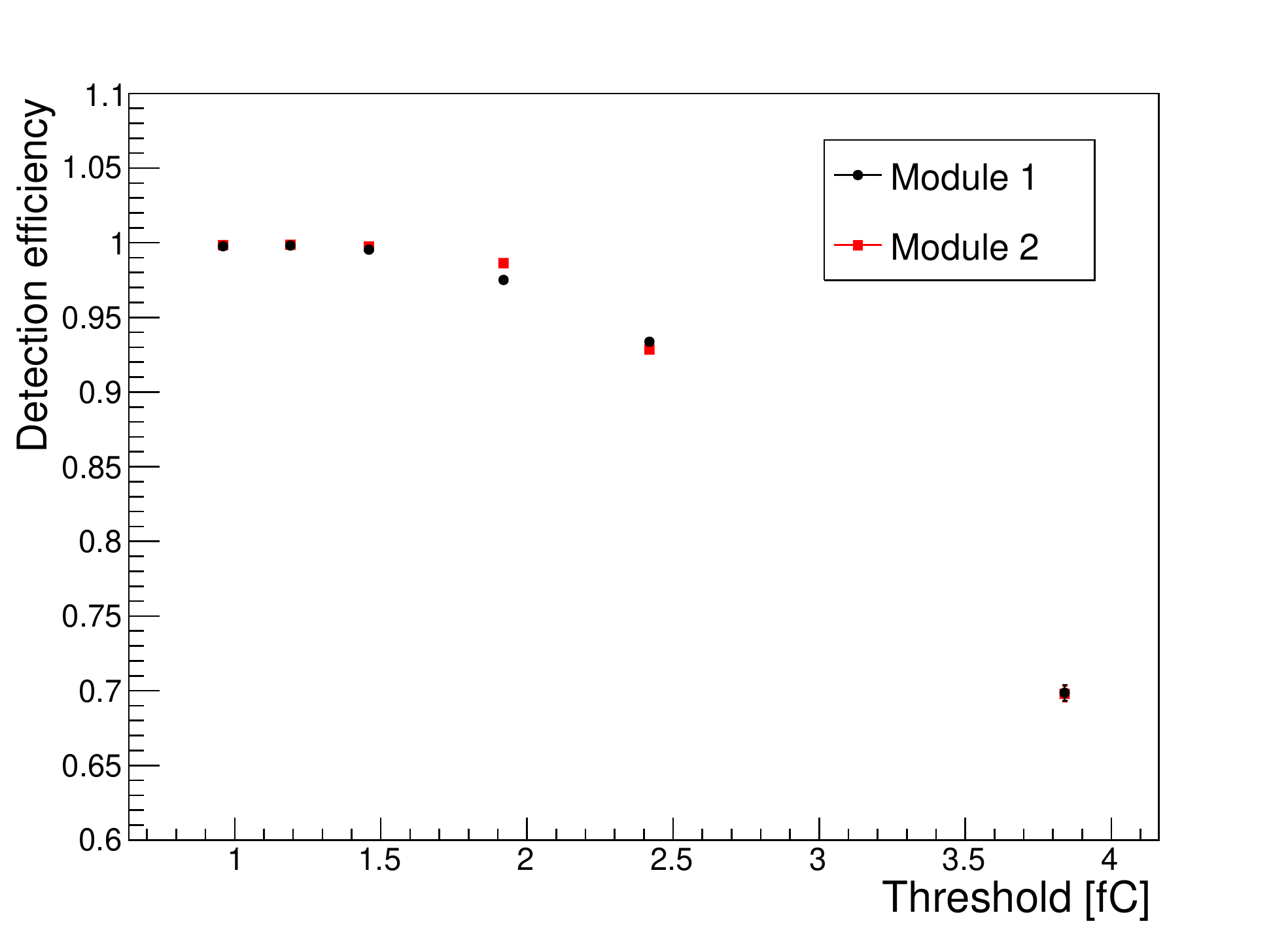}
    \caption{The detection efficiency of two detector modules as a function of the threshold.}
    \label{fig:beamtest_eff}
  \end{center}
\end{figure}

The detection efficiency is calculated by using events
with hits on two layers of the BPM and no dead sensor 
strips or no strips connected to dead readout channels
in the beam position determined by the BPM.
The detection efficiency of a module is defined as the ratio of the number of 
events with more than one hit on two modules to the number of events with more than one hit on 
the other side of module.
The measured detection efficiency is shown in figure~\ref{fig:beamtest_eff} as a function of the threshold.
The detection efficiency at a threshold of  $1.19$~fC (about 30\% of a MIP charge)
is measured to be $99.8 \pm 0.1$\%.

\begin{figure}[htbp]
  \begin{center}
    \includegraphics[width=0.49\textwidth]{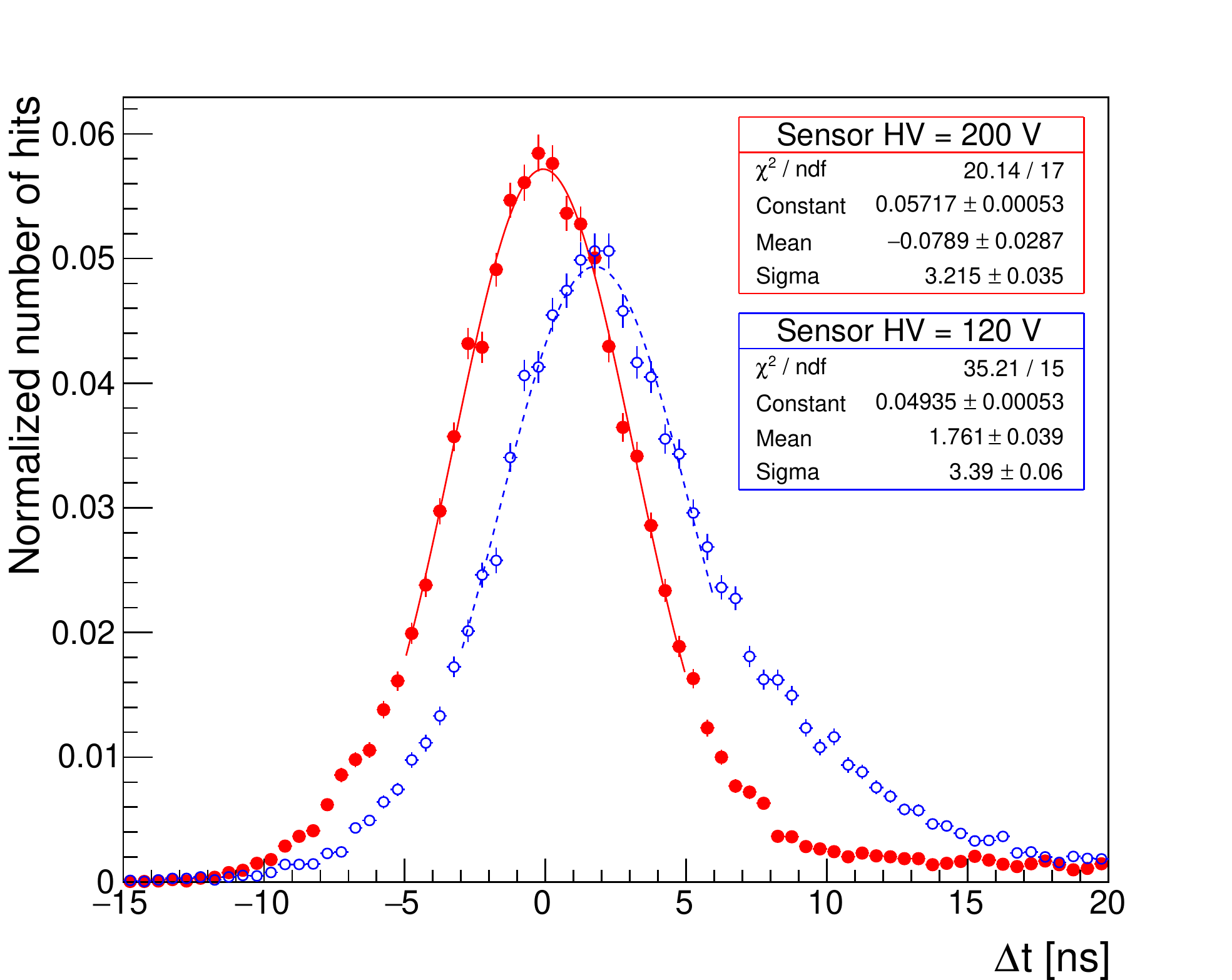}
    \includegraphics[width=0.49\textwidth]{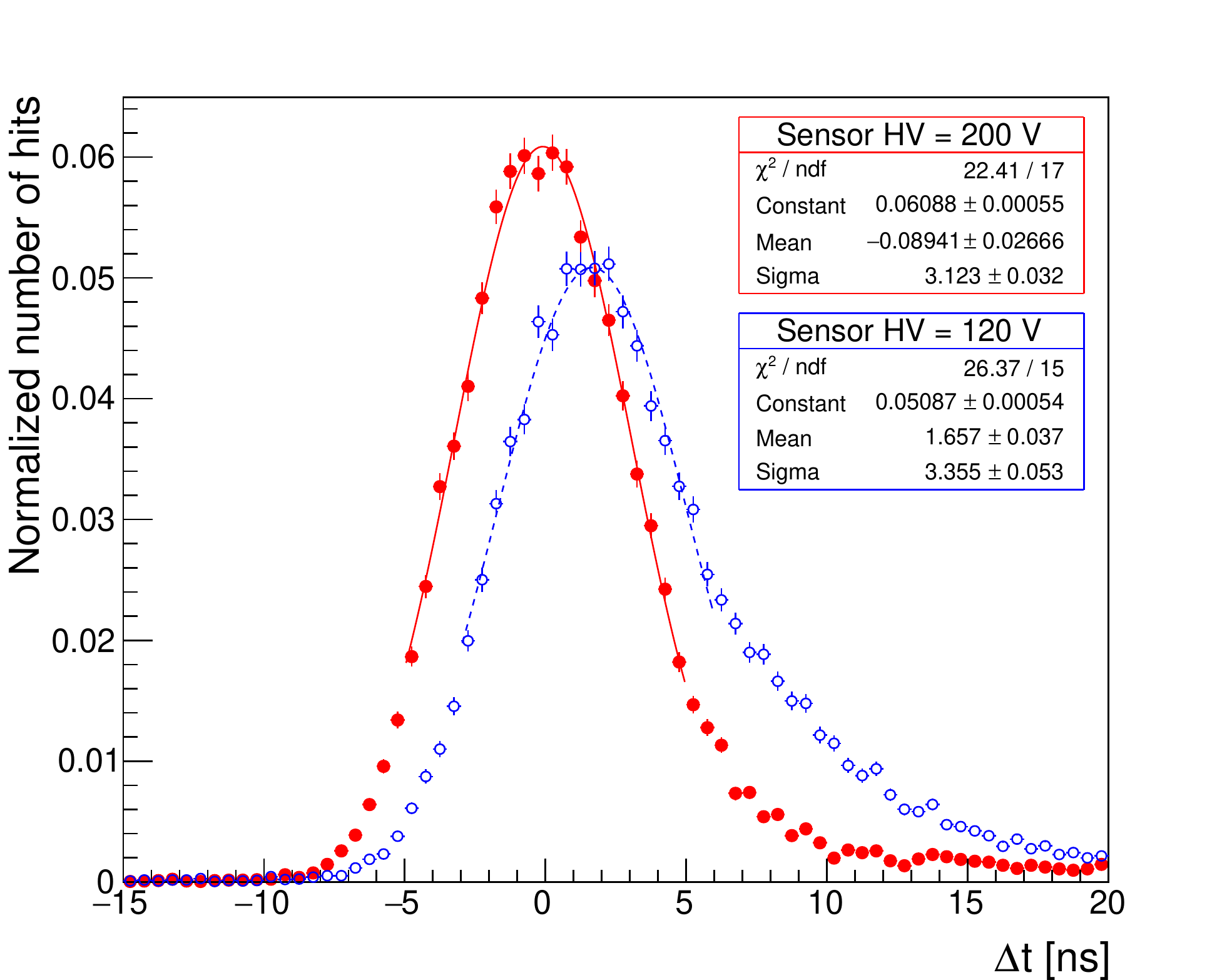}
    \caption{Distributions of the time difference ($\Delta t$) between the detector modules and trigger 
    for the Module~1 (left) and the Module~2 (right).
    The detector modules are operated with the bias voltage of 120~V (solid red circles with a solid curve) 
    and 200~V (open blue circles with a dashed curve).
    The timing correction is applied based on the data with test pulses.
    The $\Delta t$ value is shifted to adjust the peak position of the distribution at the bias voltage of 200 V to 0~ns.
    A Gaussian function is fitted to data in a core region.
}
    \label{fig:time_resolution}
  \end{center}
\end{figure}

The time resolution of the detector modules is evaluated by comparing time of 
the signal exceeding the threshold digitized in 5~ns intervals with the trigger timing.
To improve the time resolution, the timing calibration is applied based on the data with test pulses 
shown in figure~\ref{fig:TimeWalk}.
This calibration removes the time walk effect and the time difference between the channels.
To study dependence of the time resolution on the bias voltage to a sensor,
the detector modules are operated at the bias voltage of 120~V and 200~V.
The timing distributions are shown in figure~\ref{fig:time_resolution}.
The time resolution is measured to be 3.4~ns (the Module 1 and 2) as the standard
deviation in the core region 
with a long tail distribution at the bias voltage of 120~V. At the bias voltage of 200~V,
the time resolution in the core region is improved to be 3.2~ns and 3.1~ns as the standard
deviation for the Module 1 and the Module 2, respectively,
and there is a small distribution in the tail region.

\section{Conclusions}
\label{sec:conclusions}

The detector for positrons/electrons from a pulsed
muon beam was developed aiming at the measurements of the muon $g-2$/EDM 
and the muonium hyperfine structure interval at the J-PARC,
and the measurement of the proton charge radius
at Tohoku University.
The detector is required to measure a signal charge up to four MIPs,
to tolerate the hit rate of 150~kHz/mm$^{2}$, 
to have the signal-to-noise ratio greater than 15 and time resolution much better than
2~$\mu$s while it needs to store data in a period more than 10~$\mu$s.
The time walk over a signal range of 0.5-3 MIP charge is required to be
less than 1~ns to constrain a bias on the time measurement.

The detector uses a silicon strip sensor with a 190~$\mu$m strip pitch
and a front-end electronics with a sampling rate of 200~MHz and a buffer memory having a depth of 8192 points.
The data of one detector module in a period of 40.96~$\mu$s (=8192~points $\times$ 5~ns) can be read out up to a repetition frequency of 75~Hz 
without the zero-suppression
algorithm, which is sufficient for the muon beam rate of 25~Hz at the J-PARC.
Performance of fabricated detector modules was evaluated
at a laboratory and a beam test using the positron beam at Tohoku University.
According to the measured mean value of the ToT, detectors are estimated to
tolerate up to a rate of $\sim$5~MHz for a MIP charge signal while the
maximum hit rate is estimated to be 1.4~MHz for the strip size of this sensor.
The signal-to-noise ratio is estimated to be 20.8 on average for a MIP charge signal
at the actual detector for the J-PARC muon $g-2$/EDM experiment.
The detection efficiency is measured to be $99.8\pm 0.1$\% at a threshold
of 1.19~fC. 
The time resolution is measured to be 3.2~ns as the standard deviation in
the core of the distribution when the bias voltage
to a sensor is 200~V. These results satisfy the requirements of the detector.
The time walk is measured to be 17.2~ns on average and it is larger than
the requirement. This will be solved by the next version of the front-end ASIC.


\acknowledgments

The authors would like to thank the KEK and the J-PARC muon section
staffs for their strong support, the Open Source Consortium of
Instrumentation (Open-It) of KEK for their support on the electronics
design, and Simon Eidelman at Budker Institute of Nuclear Physics and
Novosibirsk State University for his diligent proofreading of this article.
This work is supported by JSPS KAKENHI Grants No. JP15H05742 and
JP16H06340.

\bibliographystyle{JHEP}
\bibliography{ms}

\providecommand{\href}[2]{#2}\begingroup\raggedright\begin{thebibliography}{10}

\bibitem{MuSE}
Y.~Miyake et~al., \emph{{Current status of the J-PARC muon facility, MUSE}},
  \href{http://dx.doi.org/10.1088/1742-6596/551/1/012061}{\emph{J. Phys.: Conf.
  Ser.} {\bfseries 551} (2014) 012061}.

\bibitem{H-line}
N.~Kawamura et~al., \emph{New concept for a large-acceptance general-purpose
  muon beamline}, \href{http://dx.doi.org/10.1093/ptep/pty116}{\emph{Prog.
  Thoer. Exp. Phys.} {\bfseries 2018} (2018) 113G01}.

\bibitem{E34_PTEP2019}
M.~Abe et~al., \emph{A new approach for measuring the muon anomalous magnetic
  moment and electric dipole moment},
  \href{http://dx.doi.org/10.1093/ptep/ptz030}{\emph{Prog. Theor. Exp. Phys.}
  {\bfseries 2019} (2019) 053C02}.

\bibitem{E821_PRD2006}
{G. W. Bennett et al. (Muon ($g-2$) Collaboration)}, \emph{{Final report of the
  E821 muon anomalous magnetic moment measurement at BNL}},
  \href{http://dx.doi.org/10.1103/PhysRevD.73.072003}{\emph{Phys. Rev. D}
  {\bfseries 73} (2006) 072003}.

\bibitem{E989_arxiv2015}
{J. Grange et al. (E989 Collaboration)}, \emph{{Muon ($g-2$) Technical Desgin
  Report}},  \href{https://arxiv.org/abs/arXiv:1501.06858}{{\ttfamily
  arXiv:1501.06858}}.

\bibitem{MuSEUM}
K.~Shimomura, \emph{{Muonium in J-PARC; from fundamental to application}},
  \href{http://dx.doi.org/10.1007/s10751-015-1159-3}{\emph{Hyperfine Interact.}
  {\bfseries 233} (2015) 89--95}.

\bibitem{ULQ2}
T.~Suda et~al., \emph{{Measurement of Proton Charge Radius by Low-Energy
  Electron Scattering}}, {\emph{J. Particle Accelerator Society of Japan}
  {\bfseries 15} (2018) 52--59}.

\bibitem{SiTCP}
T.~Uchida, \emph{{Hadware-Based TCP Processor for Gibabit Ethernet}},
  \href{http://dx.doi.org/10.1109/TNS.2008.920264}{\emph{IEEE Transactions on
  Nuclear Science} {\bfseries 55} (2008) 1631--1637}.

\bibitem{DAQMiddleware}
Y.~Yasu et~al., \emph{{A Data Acquisition Middleware}},  in \emph{2007 15th
  IEEE-NPSS Real-Time Conference}, (Batavia, IL, USA), pp.~1--3, 2007.
\newblock \href{http://dx.doi.org/10.1109/RTC.2007.4382850}{DOI}.

\bibitem{DAQMiddleware2}
Y.~Yasu et~al., \emph{{Development of DAQ-Middleware}},
  \href{http://dx.doi.org/10.1088/1742-6596/219/2/022025}{\emph{J. Phys. Conf.
  Ser.} {\bfseries 219} (2010) 022025}.

\bibitem{Ishikawa}
T.~Ishikawa et~al., \emph{{A detailed test of a BSO calorimeter with 100-800
  MeV positrons}},
  \href{http://dx.doi.org/10.1016/j.nima.2012.08.085}{\emph{Nucl. Instrum. and
  Meth. A} {\bfseries 694} (2012) 348--360}.

\end{thebibliography}\endgroup

\end{document}